\documentclass[aps,pra,superscriptaddress,showpacs,showkeys,a4paper,superscriptaddress,10pt,notitlepage,twocolumn]{revtex4-1}

\usepackage[utf8]{inputenc}
\usepackage[english]{babel}
\usepackage{amsmath}
\usepackage{amssymb}
\usepackage{amsfonts}
\usepackage{graphicx}
\usepackage{bm}
\usepackage{cancel}
\usepackage{upgreek}
\usepackage{mathtools}
\usepackage[T1]{fontenc}

\usepackage{xcolor}
\definecolor{link_blue}{RGB}{52,46,157}

\usepackage{longtable}
\usepackage{cellspace}
\usepackage[colorlinks,citecolor=link_blue,linkcolor=link_blue,urlcolor=link_blue]{hyperref}
\usepackage[tiny,center,uppercase]{titlesec}

\newcommand{\opa}[1]{\mathsf{#1}}
\newcommand\ri{\mathrm{i}}

\DeclareMathOperator{\re}{Re}

\begin{document}
	
\title{Analytic approximation for eigenvalues of a class of $\mathcal{PT}$ symmetric Hamiltonians}

\author{O.\ D.\ Skoromnik}
\email[]{olegskor@gmail.com}
\affiliation{Max Planck Institute for Nuclear Physics, Saupfercheckweg 1, 69117 Heidelberg, Germany}

\author{I.\ D.\ Feranchuk}
\email[Corresponding author: ]{ilya.feranchuk@tdt.edu.vn}
\affiliation{Atomic Molecular and Optical Physics Research Group, Ton Duc Thang University, 19 Nguyen Huu Tho Str., Tan Phong Ward, District 7, Ho Chi Minh City, Vietnam}
\affiliation{Faculty of Applied Sciences, Ton Duc Thang University, 19 Nguyen Huu Tho Str., Tan Phong Ward, District 7, Ho Chi Minh City, Vietnam}
\affiliation{Belarusian State University, 4 Nezavisimosty Ave., 220030, Minsk, Belarus}

\begin{abstract}
  An analytical approximation for the eigenvalues of $\mathcal{PT}$ symmetric Hamiltonian $\opa H = -d^{2}/dx^{2} - (\ri x)^{\epsilon+2}$, $\epsilon > -1$ is developed via simple basis sets of harmonic-oscillator wave functions with variable frequencies and equilibrium positions. We demonstrate that our approximation provides high accuracy for any given energy level for all values of $\epsilon > -1$.
\end{abstract}

\maketitle

\textit{Introduction.} --- Bender and Boettcher introduced \cite{PhysRevLett.80.5243} a family of $\mathcal{PT}$ symmetric Hamiltonians
\begin{align}
  \opa{H} = -\frac{d^{2}}{dx^{2}} - (\ri x)^{\epsilon+2}, \quad \epsilon > -1, \label{eq:1}
\end{align}
which, despite being non-Hermitian, can nevertheless possess real and discrete eigenvalues, if one considers the solution of the corresponding Schr\"{o}dinger equation on the complex plane instead of the real axis \cite{BENDER1993442,PhysRev.184.1231}. This remarkable discovery triggered theoretical investigations \cite{0034-4885-70-6-R03,1751-8121-40-32-R01,0305-4470-34-28-305,0305-4470-33-1-101,0305-4470-33-1-101,0305-4470-33-48-314,ZNOJIL1999108,0305-4470-32-17-303,Fring20172318,PhysRevA.95.052128,PhysRevA.95.053626,PhysRevA.95.053613,PhysRevA.95.053830,PhysRevLett.118.056401,PhysRevLett.118.130201,1367-2630-18-6-065005,1751-8121-49-21-215304,MIAO20161805,PhysRevA.93.031802,PhysRevA.93.012123,1751-8121-49-10-10LT03,1751-8121-50-3-035601,1751-8121-49-45-45LT01,PhysRevA.96.011802,PhysRevLett.104.061601,PhysRevA.96.012127,PhysRevA.96.013845,PhysRevLett.119.033905,0305-4470-38-1-013,PhysRevLett.101.080402} and experimental activities towards the realization of systems that can be effectively described by $\mathcal{PT}$ symmetric Hamiltonians \cite{bo_2014_5_394,xiao2017observation,yuto2017parity,wimmer2015observation,PhysRevA.88.062111,Peng328,PhysRevA.84.040101,Zheng20120053,PhysRevLett.109.150405,Regensburger488_167_2012,PhysRevLett.108.173901,PhysRevA.84.021806,PhysRevLett.106.093902,PhysRevLett.108.024101,PhysRevLett.99.167003,Feng729,PhysRevA.81.042903,Ruter_Nature_6_192_2010}.

Usually, the numerical solution of the Schr\"{o}dinger equation with the Hamiltonian (\ref{eq:1}) is performed either via the shooting method \cite{PhysRevLett.80.5243,doi:10.1063/1.532860}, basis expansion \cite{doi:10.1063/1.532860}, by discretizing the Schr\"{o}dinger equation and applying the Arnoldi iteration \cite{PhysRevA.95.052113}, or via the solution of an integral equation \cite{1751-8121-40-32-R01}.

On the other hand, analytical approximations can provide qualitative peculiarities of the system in a broad range of quantum numbers and parameters of the Hamiltonian and can serve as a basis for further numerical solutions. Moreover, the applicability of the known approximations for the Hermitian systems can be investigated in the non-Hermitian case. As such, an analytical approximation for the Hamiltonian (\ref{eq:1}) when $\epsilon \ge 0 $ was derived through the modified quasi-classical WKB approximation \cite{PhysRevLett.80.5243}, when the turning points are located on the complex plane.

In addition, the energy levels were determined by employing the variational method \cite{BENDER1999224} for the three-parameter functional $\langle \opa H(a,b,c)\rangle = \int_{C}dx \psi(x) \opa H \psi(x) / \int_{C}\psi^{2}(x)dx$, where the path $C$ on the complex plane is chosen in such a way that the trial function $\psi(x) = (\ri x)^{c} \exp(a (\ri x)^{b})$ is exponentially decaying at infinity. While this procedure straightforwardly applies to the ground state, the calculation of the energies of the excited states is difficult due to the necessity of introducing the supersymmetric partner for the Hamiltonian. This requires the evaluation of an increasing number of derivatives, which in turns demands the knowledge of the ground state wave function with very high accuracy.

In our work we propose a simple approach for the evaluation of an energy level for any given quantum number $n$ and all values of the parameter $\epsilon > -1$ with high accuracy.
\begin{figure}[t]
  \centering
  \includegraphics[width=\columnwidth]{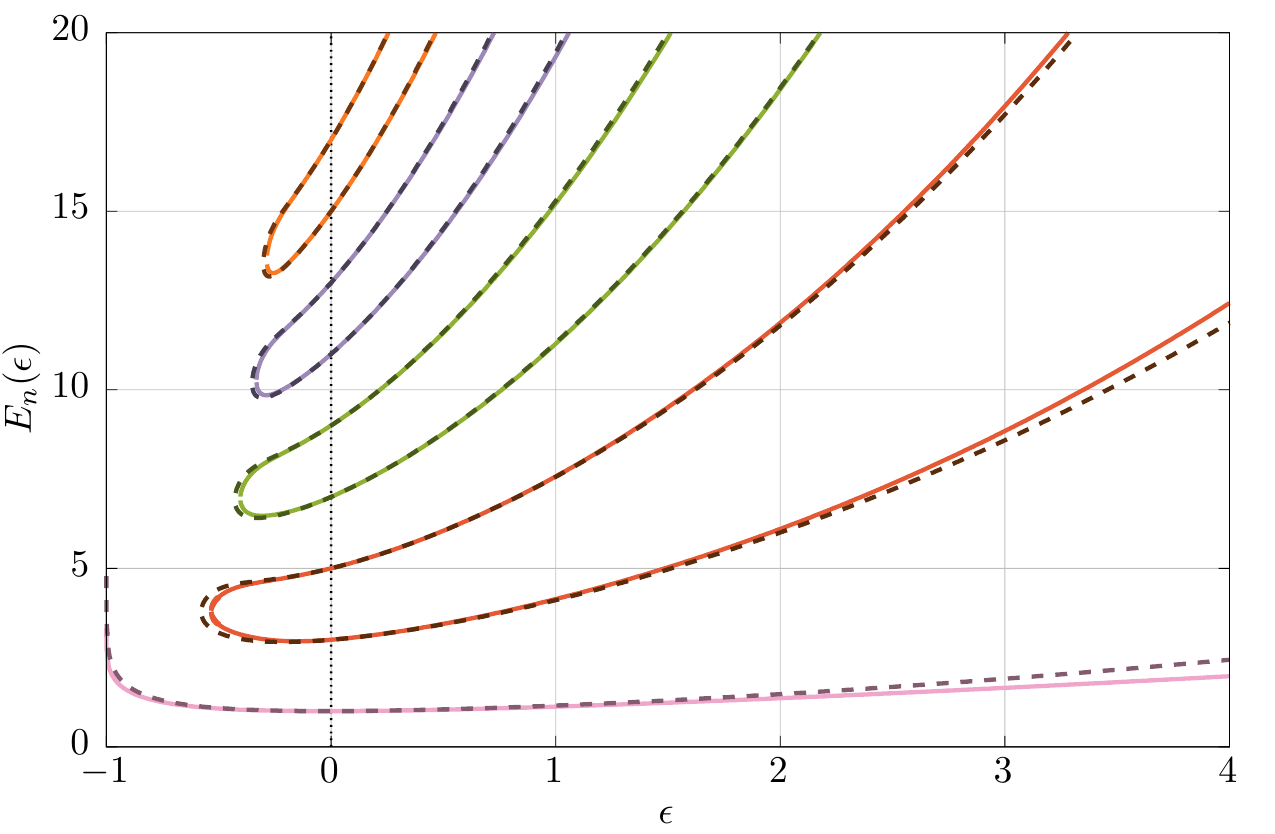}
  \caption{(Color online) The dependence of the eigenvalues $E_{n}$ on the parameter $\epsilon$. The dashed lines are the exact numerical solutions via the Arnoldi iteration. The solid lines are the analytic approximation.}
  \label{fig:1}
\end{figure}

The idea is based on the method \cite{Feranchuk2015,Feranchuk1995370} which was employed for the nonperturbative description of various Hermitian quantum systems. Consequently, we firstly illustrate the idea in the Hermitian case.

\textit{Method description.} --- Let us introduce sets $\nu_{\alpha}^{i} \equiv \nu_{\alpha}^{1}, \nu_{\alpha}^{2},\ldots, \nu_{\alpha}^{s}$, $\alpha = 0, 1, \ldots$ of $s$ parameters and a complete basis $|\psi_{n}(\nu_{\alpha}^{i})\rangle$, which depends on these parameters. The index $n$ here numerates the state vectors. For any given set of $s$ parameters $\nu_{\alpha}^{i}$ the basis functions are orthonormal $\langle\psi_{n}(\nu_{\alpha}^{i})|\psi_{k}(\nu_{\alpha}^{i})\rangle = \delta_{nk}$. However, for two different sets of parameters $\nu_{\alpha}^{i}$ and $\nu_{\alpha'}^{i}$, $\alpha \neq \alpha'$ the basis functions are not orthogonal $\langle\psi_{n}(\nu_{\alpha}^{i})|\psi_{k}(\nu_{\alpha'}^{i})\rangle = S^{\alpha\alpha'}_{nk}$, i.e. the overlapping integral $S^{\alpha\alpha'}_{nk}$ is non-vanishing.

For example, let us consider harmonic oscillator wave functions $y_{n}(x)$. Then the notations $y_{n}(\{\omega_{1},u_{1}\}, x + u_{1})$ and $y_{n}(\{\omega_{2}, u_{2}\}, x + u_{2})$ mean that we have two sets of harmonic oscillator wave functions, numbered by the same index $n = 0, 1, 2, \ldots$ which, however, have two different frequencies and two different equilibrium positions. Therefore in this case a set of parameters $\{\nu_{1}^{1},\nu_{1}^{2}\}$ ($\{\nu_{2}^{1},\nu_{2}^{2}\}$) is given by the set $\{\omega_{1}, u_{1}\}$ ($\{\omega_{2}, u_{2}\}$) correspondingly. Consequently, the basis functions with the same $\{\omega,u\}$ are orthogonal to each other
\begin{align*}
  \int_{-\infty}^{\infty}y_{n}(\{\omega_{1},u_{1}\}, x + u_{1})y_{k}(\{\omega_{1},u_{1}\}, x + u_{1}) dx = \delta_{nk},
\end{align*}
while for different $\{\omega,u\}$ the functions $y_{n}(\{\omega_{1},u_{1}\},x + u_{1})$ and $y_{k}(\{\omega_{2},u_{2}\}, x + u_{2})$ are not orthogonal and define an overlapping integral $S_{nk}^{12}$. Another example can include hydrogen like basis sets with different charges and localization of nuclei.

Suppose that the eigenvalues $E_{n}$ and eigenvectors $|\Psi_{n}\rangle$  of the stationary Schr\"{o}dinger equation need to be found: $\opa H |\Psi_{n}\rangle = E_{n}|\Psi_{n}\rangle$, $n = 0,1,\ldots$. Let us take the state vector $|\Psi_{n}\rangle$ with the corresponding eigenvalue $E_{n}$ and expand this state vector in a complete basis $|\psi_{k}(\nu_{n}^{i})\rangle$
\begin{align}
  |\Psi_{n}\rangle = |\psi_{n}(\nu_{n}^{i})\rangle + \sum_{k \neq n} C_{n k}|\psi_{k}(\nu_{n}^{i} )\rangle.\label{eq:2}
\end{align}
Here we set the index $\alpha = n$ in order to highlight that for different state vectors $|\Psi_{n}\rangle$ the parameters can be different. In addition, the state vector $|\Psi_{n}\rangle$ is normalized as $\langle\Psi_{n}|\psi_{n}(\nu_{n}^{i})\rangle = 1$. By plugging Eq.~(\ref{eq:2}) into the Sch\"{o}dinger equation and projecting on the state vector $|\psi_{n}(\nu_{n}^{i})\rangle$ and $|\psi_{l}(\nu_{n}^{i})\rangle$, $l \neq n$ one obtains the system of nonlinear equations for the determination of the energy $E_{n}$ and coefficients $C_{n k}$, $k\neq n$:
\begin{widetext}
  \begin{align}
    E_{n} &= H_{n n}(\nu_{n}^{i} ) + \sum_{k \neq n} C_{n k}H_{n k}(\nu_{n}^{i} ), \label{eq:3}
    \\
    C_{n k} &= \left[E_{n} - H_{kk}(\nu_{n}^{i} )\right]^{-1}\left[ H_{k n}(\nu_{n}^{i} ) + \sum_{l \neq n \neq k}C_{n l} H_{k l}(\nu_{n}^{i} )\right], \quad k \neq n, \label{eq:4}
  \end{align}  
\end{widetext}
where $H_{n k}(\nu_{n}^{i} ) \equiv \langle\psi_{n} (\nu_{n}^{i} )| \opa H | \psi_{k} (\nu_{n}^{i})\rangle$.

We stress here that the system of equations (\ref{eq:3}), (\ref{eq:4}) is completely equivalent to the original Schr\"{o}dinger equation. Moreover, the system of equations for the determination of the energy level $E_{n}$ is completely decoupled from the system of equations for the energy level $E_{n'}$, $n \neq n'$. This should not be confused with the Galerkin method \cite{GalerkinMethod}, where the Hamiltonian of the system is written in a certain basis and the energy levels are determined as all eigenvalues of the system of $N$ linear equations $\sum_{k}(H_{nk} - E_{n}\delta_{nk})C_{nk} = 0$. Instead in our case we have the system of nonlinear equations for any given energy level and in order to determine $N$ eigenvalues one needs to solve $N$ nonlinear systems of equations.

By construction of the system of equations (\ref{eq:3}), (\ref{eq:4}) we can employ the different parameters $\nu_{n}^{i}$ and $\nu_{n'}^{i}$ when seeking different eigenvalues $E_{n}$ and $E_{n'}$. This of course requires the recalculation of the matrix elements. For example, imagine that the Hamiltonian in a certain single-parametric basis is represented as a $3\times3$ matrix. In order to determine the energy level $E_{0}$ one employs parameter $\nu_{0}$ for the calculation of the matrix elements and obtains the system of equations for three unknowns, namely $E_{0}$, $C_{01}$ and $C_{02}$. However, in order to calculate the energy level $E_{1}$ one can employ a different parameter $\nu_{1} \neq \nu_{0}$ and obtain a system of equations for $E_{1}$, $C_{10}$ and $C_{12}$ in which the matrix element $H_{10}(\nu_{1}) \neq H_{01}^{*}(\nu_{0})$, since $\nu_{1}\neq \nu_{0}$.

In addition, we mention here that if one can solve the system of Eqs.~(\ref{eq:3}), (\ref{eq:4}) exactly then the choice of the parameters is not important. In practice, however, it is either very complicated or impossible to solve the system of Eqs.~(\ref{eq:3}), (\ref{eq:4}) exactly. Consequently, one seeks for an approximate solution. It is exactly here, when the different values of parameters come to hand and become very effective. The idea of our approach consists in adjusting the parameters in a way that the diagonal element of the Hamiltonian $H_{nn}(\nu_{n}^{i})$ becomes the best possible approximation for the eigenvalue $E_{n}$. As an example, let us examine the case when the Hamiltonian is represented as $3 \times 3$ matrix in a basis which depends on two parameters $\nu_{n}^{1}$ and $\nu_{n}^{2}$, $n = 0, 1, 2$. If one fixes the parameters $\nu_{0}^{1}$ and $\nu_{0}^{2}$ from the system of equations
\begin{align*}
  H_{01}(\nu_{0}^{1}, \nu_{0}^{2}) = 0,\quad  H_{02}(\nu_{0}^{1}, \nu_{0}^{2}) = 0,
\end{align*}
then the energy $E_{0}$ is simply given as the diagonal matrix element $H_{00}(\nu_{0}^{1}, \nu_{0}^{2})$. Consequently, one can repeat this procedure for $E_{1}$ and $E_{2}$ and obtain these energy levels as the corresponding diagonal entities of the Hamiltonian $H_{11}(\nu_{1}^{1}, \nu_{1}^{2})$, $H_{22}(\nu_{2}^{1}, \nu_{2}^{2})$ calculated with the corresponding parameters, found from the vanishing matrix elements.

In reality, the Hamiltonian of the system in an $s$-parametric basis is represented via an infinitely dimensional matrix and consequently one can make equal to zero only $s$ matrix elements. The remaining contributions due to the non-vanishing matrix elements should be taken into account via Eqs.~(\ref{eq:3}), (\ref{eq:4}). However, if the diagonal entity of the Hamiltonian $H_{nn}(\nu_{n}^{i})$ with this choice of parameters provides a good approximation for the exact eigenvalue $E_{n}$, then the corrections due to the off-diagonal elements can be taken into account perturbatively, via the iteration scheme (see Refs.~\cite{Feranchuk2015,Feranchuk1995370}).

The main advantage of the above described approach is in fact that if the matrix elements can be evaluated analytically, then instead of the solution of the differential equation one employs a root search algorithm for the determination of the values of the parameters. This procedure was successfully applied to a variety of problems with Hermitian Hamiltonians, for example, one dimensional problems \cite{Feranchuk2015} including the anharmonic oscillator \cite{Feranchuk1995370}, the quantum Rabi model \cite{1751-8121-49-45-454001,*0305-4470-29-14-026}, the polaron problem \cite{0022-3719-17-24-012} and scalar quantum field theory \cite{PhysRevD.92.125019}. In addition it was recently employed for multi-electron atoms \cite{1701.04800}, which provided an analytic approximation for their energy levels with a relative accuracy of $5\%$ for the whole periodic table.

\textit{The quantum anharmonic oscillator example.} --- For example, for the quantum anharmonic oscillator with the Hamiltonian: 
\begin{align}
  \opa H = -\frac{1}{2}\frac{d^{2}}{dx^{2}} + \frac{1}{2} x^{2} + \lambda x^{4}, \quad \lambda > 0, \label{eq:5}
\end{align}
the approximate solution can be found completely analytically. One employs the single-parametric harmonic oscillator wave functions $y_{n}(\omega_{n}, x)$ with the frequency $\omega_{n}$ and finds non-vanishing matrix elements
\begin{widetext}
  \begin{equation}
    \begin{aligned}
      H_{nn}(\omega_{n}) &= \frac{1}{4\omega_{n}}(\omega_{n}^{2}+1)(1+2n)+\frac{3\lambda}{4\omega_{n}^{2}}(1+2n+2n^{2}),
      \\
      H_{nk}(\omega_{n}) &= H_{kn}(\omega_{n}) = \frac{1}{4}\sqrt{(n+1)(n+2)}\left(\frac{1-\omega_{n}^{2}}{\omega_{n}} + \frac{2\lambda}{\omega_{n}^{2}}(2n+3)\right)\delta_{n+2,k} + \frac{\lambda}{4\omega_{n}^{2}}\sqrt{\frac{(n+4)!}{n!}}\delta_{n+4,k}.      
    \end{aligned}\label{eq:6}
  \end{equation}  
\end{widetext}
Since we introduced the single-parametric basis, we can make equal to zero only a single matrix element. The closest matrix element for a given state $n$ is $H_{n,n+2}$. Consequently, from the condition $H_{n,n+2} = 0$ one trivially finds the equation for $\omega_{n}$
\begin{align}
  \omega_{n}^{3} - \omega_{n} - 2\lambda(2n + 3) = 0.\label{eq:7}
\end{align}
From here, the energy levels of the anharmonic oscillator are defined through $E_{n} = H_{nn}(\omega_{n})$, where $\omega_{n}$ is defined by Eq.~(\ref{eq:7}). Interesting enough this provides a uniform approximation for different $n$ and $\lambda$. Compare the exact solution with this simple analytical result (see Table \ref{tab:1}).
\begin{table}[h]
  \centering
  \begin{tabular}{| c | c  c  c  c|}
    \hline \hline
     & $\lambda$ & & &\\
     $E_{n}^{\mathrm{Analytic}}$ ($E_{n}^{\mathrm{Exact}}$) & 0.1 & 1 & 10 & 100\\
    \hline
    $n = 0$ & 0.5603 & 0.8125 & 1.5313 & 3.1924 \\
            & (0.5591) & (0.8038) & (1.5050) & (3.1314) \\
    $n = 10$ & 17.3748 & 32.9931  & 68.9367 & 147.515 \\
             & (17.3519) & (32.9333) & (68.8037) & (147.227) \\
    $n = 40$ & 96.0745 & 195.865  & 416.735  & 895.387 \\
             & (95.5602) & (194.602) & (413.938) & (889.325) \\
     \hline \hline
  \end{tabular}
  \caption{Comparison of the eigenvalues for the anharmonic oscillator with the Hamiltonian (\ref{eq:5}) obtained through the analytic approximation with the ones calculated numerically via the Arnoldi iteration.}
  \label{tab:1}
\end{table}

In addition, we mention here that the standard perturbation theory series for any $\lambda \neq 0$ has zero radius of convergence \cite{PhysRev.184.1231}.

\textit{Application to the family of $\mathcal{PT}$-symmetric Hamiltonians (\ref{eq:1}).} --- Here, we apply this approach for the determination of an analytic approximation for the eigenvalues of non-Hermitian Hamiltonian (\ref{eq:1}), following the steps outlined below.

First, we introduce a basis. Since for $\epsilon = 0$ the exact solution of the problem is known, it is quite natural to employ a complete and orthogonal harmonic oscillator basis, which has a single quantum number $n$ and frequency $\omega$. It is well known that in order to perform the exact diagonalization of the Hamiltonian $\opa H = -d^{2}/dx^{2} + x^{2} + x$ one needs to shift the equilibrium position of the harmonic oscillator. In addition, for $\epsilon = 2$ the Hamiltonian (\ref{eq:1}) is equivalent \cite{0034-4885-70-6-R03} to $\opa H = -d^{2}/dx^{2} + 4x^{4} - 2x$, which has also linear $x$ term. For this reason, we assume that our basis should have shifted equilibrium position. Moreover, as we are dealing with the non-Hermitian Hamiltonians the equilibrium position can be shifted into the complex plane, i.e., $x \to x + \ri u$. This should not be confused with the complex scaling method \cite{MOISEYEV1998212,doi:10.1146/annurev.pc.33.100182.001255}, when the rotation of the integration contour is performed, i.e., $x\to x\exp(\ri \theta)$. Thus, we arrive to the two parametric harmonic oscillator basis
\begin{align}
  y_{n}(\omega_{n}, u_{n}, x) &= \sqrt[4]{\frac{\omega_{n}}{\pi}}\frac{1}{\sqrt{2^{n}n!}}\exp\left(-\frac{\omega_{n}}{2}(x + \ri u_{n})^{2}\right) \nonumber
  \\
  &\mspace{100mu}\times H_{n}(\sqrt{\omega_{n}}(x + \ri u_{n})). \label{eq:8}
\end{align}
Here $H_{n}(x)$ is the Hermitian polynomial and we have $\omega_{n}$ and $u_{n}$ as parameters, which can be different for different states. For this reason, we denoted them with the index $n$. Consequently, the states with different $n$ can have different $\omega_{n}$ or $u_{n}$ and can be non-orthogonal.

Second, our basis functions contain only two free parameters and therefore, by adjusting them only two matrix elements, for a given energy level, can be made equal to zero. The remaining matrix elements should be taken into account by means of the iteration scheme of Refs.~\cite{Feranchuk2015,Feranchuk1995370}. However, since we are looking for simple analytical expressions for eigenvalues we limit ourselves only to the zeroth-order approximation and disregard the contributions due to the remaining off-diagonal elements. We consider that the final answer justifies this approximation.
\begin{figure*}[t]
  \centering
  \includegraphics[width=\columnwidth]{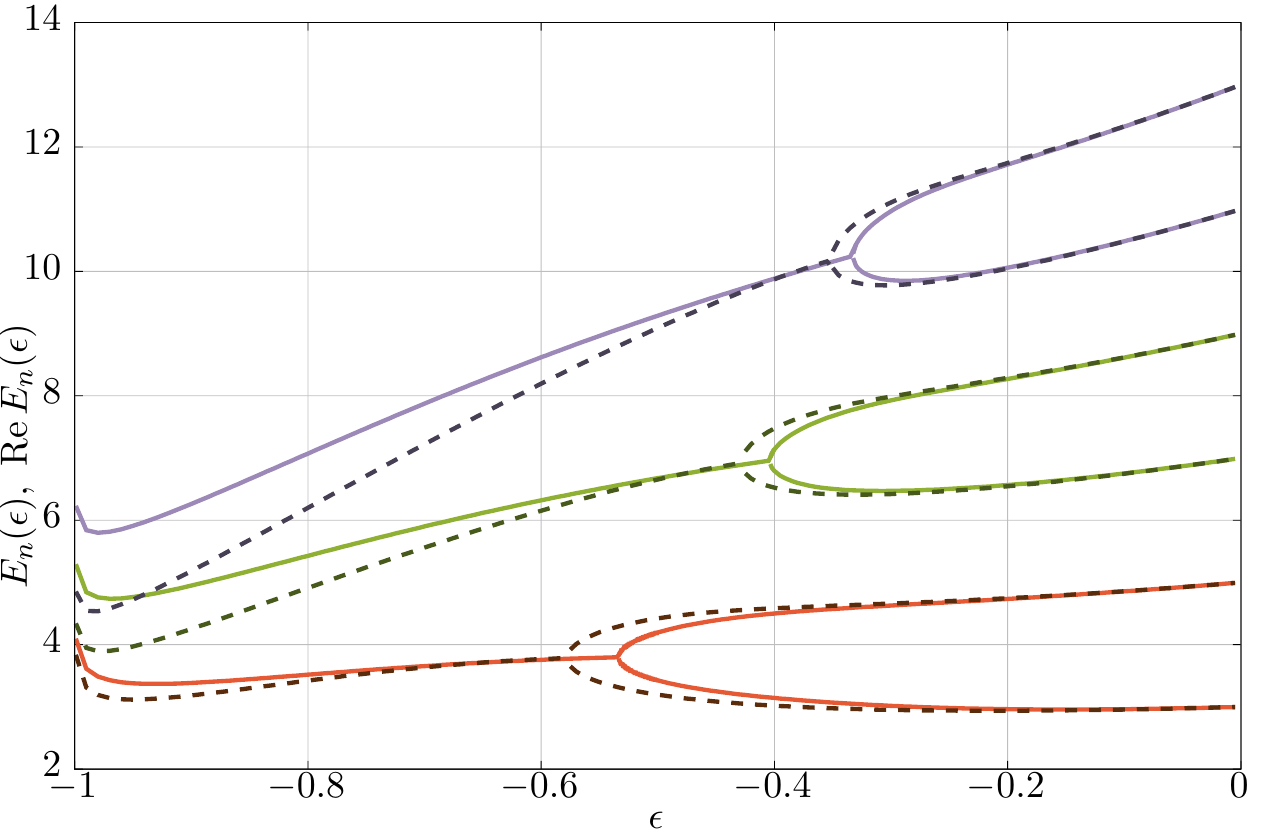}
  \includegraphics[width=\columnwidth]{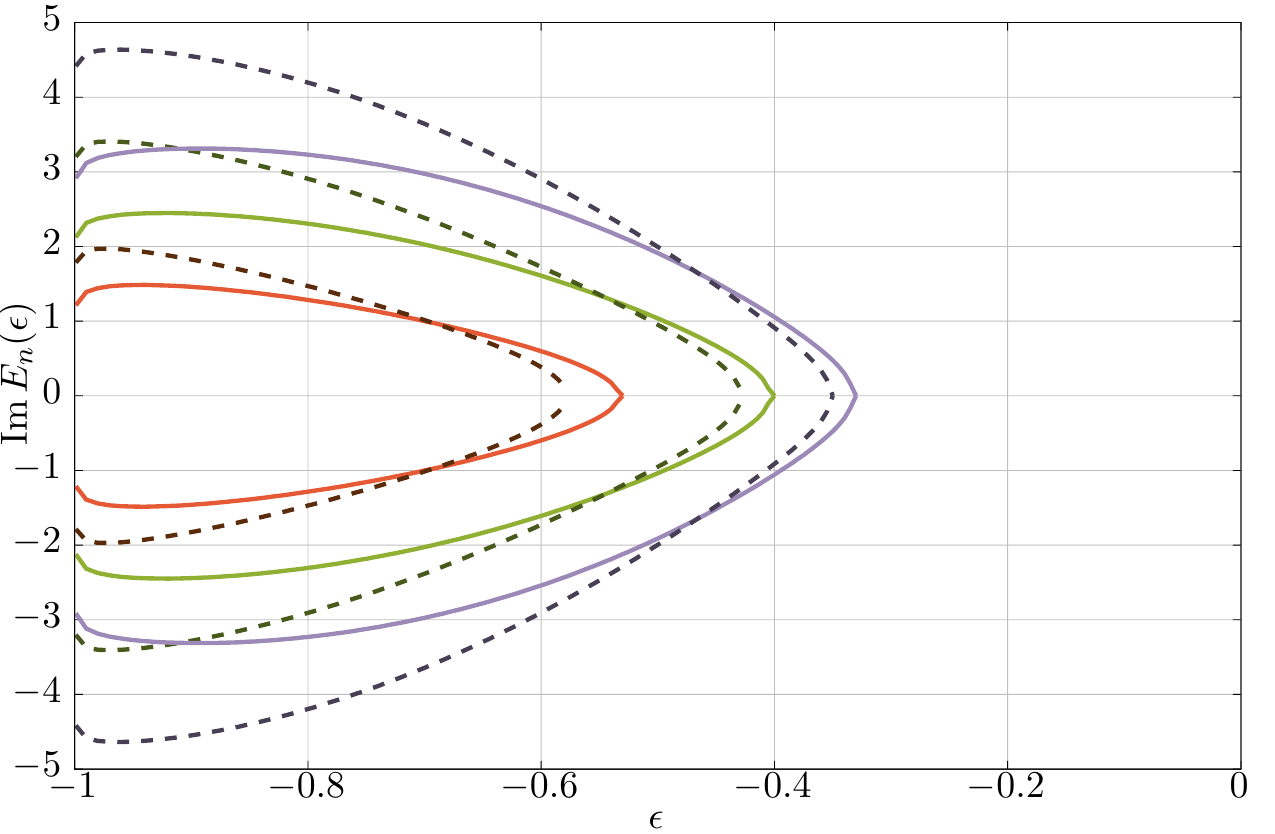}
  \caption{(Color online) The dependence of the eigenvalues $E_{n}$, their real and imaginary parts on the parameter $\epsilon$, in the region $-1 < \epsilon \le 0$. The dashed lines are the exact numerical solutions via the Arnoldi iteration. The solid lines are the analytic approximation via Eq.~(\ref{eq:19}). After the point at $\epsilon = \epsilon_{\text{branching point}}$, where two energy level coalesce they appear as complex conjugate pairs, whose real parts together with the real energy levels are plotted on the left pane and imaginary parts on the right pane respectively.}
  \label{fig:2}
\end{figure*}

Third, the scalar product for the calculation of the matrix elements in the harmonic oscillator basis, with $u_{n}$ and $\omega_{n}$ fixed and real, is defined in a $\mathcal{PT}$ symmetric way
\begin{equation}
  (y_{n},y_{m}) = \int_{-\infty}^{\infty} y_{n}^{\star}(x) y_{m}(x)dx = (-1)^{n}\delta_{nm},\label{eq:9}
\end{equation}
\begin{equation}
  y_{n}^{\star}(x) = \mathcal{PT} y_{n}(x) = y^{*}_{n}(-x) = (-1)^{n}y_{n}(x).\label{eq:10}
\end{equation}
The relation (\ref{eq:10}) follows from the fact that under operation of $\mathcal{PT}$-symmetry the argument of the function changes into $-(x + \ri u)$, that the exponent has quadratic argument and the property of the Hermitian polynomials $H_{n}(-x) = (-1)^{n} H_{n}(x)$. The relation (\ref{eq:9}) follows from Eq.~(\ref{eq:10}) and the orthogonality of harmonic oscillator wave functions.

\begin{widetext}
  Fourth, a simple evaluation of the expectation value of the Hamiltonian and the matrix elements in the harmonic oscillator basis with $u_{n}$ and $\omega_{n}$ fixed and real yields
  \begin{align}
    H_{nn}(\omega_{n},u_{n}) &= \frac{(y_{n},H y_{n})}{(y_{n},y_{n})} = \frac{1}{2}\omega_{n} (2n+1)-\frac{1}{\omega_{n}^{(\epsilon+2)/2}}\frac{1}{\sqrt{\pi}2^{n}n!}\int_{-\infty}^{\infty}e^{-(x+\ri u_{n})^{2}}H_{n}^{2}(x + \ri u_{n})(\ri x)^{\epsilon+2}dx, \label{eq:11}
    \\
    H_{nk}(\omega_{n},u_{n}) &= (-1)^{n}\Bigg(-\frac{\omega_{n}}{2}\sqrt{n(n-1)}\delta_{k,n-2} - \frac{\omega_{n}}{2}\sqrt{(n+1)(n+2)}\delta_{k,n+2} \nonumber
    \\
           &\mspace{90mu}- \frac{1}{\omega_{n}^{(\epsilon+2)/2}}\frac{1}{\sqrt{\pi}\sqrt{2^{n}n!2^{k}k!}}\int_{-\infty}^{\infty}e^{-(x+\ri u_{n})^{2}}H_{n}(x+\ri u_{n})H_{k}(x+\ri u_{n})(\ri x)^{\epsilon+2}dx\Bigg), \quad k \neq n.\label{eq:12}
  \end{align}
\end{widetext}
Moreover, the integrals in Eqs.~(\ref{eq:11}-\ref{eq:12}) can be evaluated analytically and are expressed through the parabolic cylinder special functions $D_{\nu}(x)$ \cite{whittaker1996course}. This is achieved with the help of the following integral \cite{gradshteyn2014table}
\begin{align}
  \int_{-\infty}^{\infty}&(\ri x)^{\nu} e^{-\beta^{2}x^{2} - \ri q x}dx = \nonumber
  \\
  &2^{-\frac{\nu}{2}}\sqrt{\pi}\beta^{-\nu-1}\exp\left(-\frac{q^{2}}{8\beta^{2}}\right)D_{\nu}\left(\frac{q}{\beta\sqrt{2}}\right),\label{eq:13}
\end{align}
where $\re\beta > 0$ and $\nu > -1$. For example, when $n = 0$ and $k = 1$ we get
\begin{align}
  H_{00}(\omega_{0},u_{0}) = \frac{\omega_{0}}{2} - \frac{e^{u_{0}^{2}/2} 2^{-\frac{\epsilon+2}{2}}}{\omega_{0}^{(\epsilon+2)/2}} D_{\epsilon+2}(\sqrt{2}u_{0})\label{eq:14}
\end{align}
and
\begin{align}
  H_{01}(\omega_{0},u_{0}) &= \frac{\ri e^{\frac{u_{0}^{2}}{2}}2^{-\frac{\epsilon+4}{2}}}{\omega_{0}^{(\epsilon+2)/2}}\Big(D_{\epsilon+3}(\sqrt{2}u_{0}) \nonumber
  \\
  &\mspace{120mu}- \sqrt{2}u_{0} D_{\epsilon+2}(\sqrt{2}u_{0})\Big).\label{eq:15}
\end{align}
  
\begin{widetext}
  Fifth, according to the above described calculation scheme for the Hermitian case the parameters $\omega_{n}$ and $u_{n}$ are determined from the condition that the nearest matrix elements are vanishing. Since we have only two parameters, only two matrix elements can be made equal to zero. In the case of Eqs.~(\ref{eq:11}), (\ref{eq:12}) the nearest matrix elements are $H_{n, n+1}(\omega_{n},u_{n})$ and $H_{n,n+2}(\omega_{n},u_{n})$. The solution of the system of equations
  \begin{equation}
    \left\{\begin{aligned}
        H_{n,n+1}(\omega_{n},u_{n}) = 0
        \\
        H_{n,n+2}(\omega_{n},u_{n}) = 0
      \end{aligned}\right. \label{eq:16}
  \end{equation}
  gives $\omega_{n}$ and $u_{n}$. The equation for $\omega_{n}$ can be solved explicitly which yields
  \begin{align}
    \omega_{n} = \left( -\frac{1}{\sqrt{\pi} 2^{n}n!(n+1)(n+2)}\int_{-\infty}^{\infty}e^{-(x+\ri u_{n})^{2}}H_{n}(x+\ri u_{n})H_{n+2}(x+\ri u_{n}) (\ri x)^{\epsilon+2}dx\right)^{\frac{2}{\epsilon+4}},\label{eq:17}
  \end{align}
  while the values of $u_{n}$ are determined as the roots of the function $g(u_{n})$
  \begin{align}
    g(u_{n}) = \frac{1}{\sqrt{\pi}2^{n}n!}\frac{1}{\sqrt{2(n+1)}}\int_{-\infty}^{\infty} e^{-(x+\ri u_{n})^{2}}H_{n}(x+\ri u_{n})H_{n+1}(x+\ri u_{n})(\ri x)^{\epsilon+2}dx, \quad \Rightarrow g(u_{n}) = 0\quad  \Rightarrow u_{n}, \label{eq:18}
  \end{align}
\end{widetext}
We also add here that for integer values of $\epsilon$ the roots of the function $g(u_{n})$ can be found exactly, as the parabolic cylinder functions are expressed through polynomials. This yields either quadratic or biquadratic equations for $u_{n}$. For other values of $\epsilon$ the numerical search for roots was employed.

Sixth, the previous analysis demonstrates \cite{PhysRevLett.80.5243} that when $\epsilon < -0.57793$ all energy levels except of the ground state coalesce. For this reason, in the vicinity of the exceptional points expression (\ref{eq:3}) in not applicable as is in the zeroth-order approximation and should be modified as in the perturbation theory case for the doubly degenerate levels \cite{LandauQM}. An analogous linear combinations of multiple energy levels were also employed in magnetohydrodynamics \cite{0305-4470-39-32-S08}, $\mathcal{PT}$-symmetric Bose-Hubbard model \cite{1751-8121-41-25-255206} and models of multiple coupled wave guides \cite{1751-8121-45-2-025303,1751-8121-49-49-495303}.

Consequently, we form a linear combination of odd $n=1, 3, 5,\ldots$ and even $n = 2, 4, 6, \ldots$ states instead. Being precise, we mixed $(1,2)$, $(3,4)$, $\ldots$ states. Then the energy levels are determined from the system of linear equations \cite{LandauQM}
\begin{align}
  E_{1,2} &= \frac{1}{2}(H_{11}(\omega_{n},u_{n}) + H_{22}(\omega_{n},u_{n})) \label{eq:19}
  \\
  &\pm \frac{1}{2}\sqrt{(H_{11}(\omega_{n},u_{n}) - H_{22}(\omega_{n},u_{n}))^{2} + 4H_{12}^{2}(\omega_{n},u_{n})}.\nonumber
\end{align}
Here $H_{11}$ is the expectation value for the odd states, $H_{22}$ for the even states and $H_{12}$ is the matrix element between the odd and the even states. We pay attention here that due to the $\mathcal{PT}$-symmetric definition of the scalar product the square of the matrix element but not the square of the absolute value appears under the square root in Eq.~(\ref{eq:19}). Since the matrix element between the odd and the even states in this case can not be equal to zero, we choose $u_{n}$ and $\omega_{n}$ from the vanishing $H_{(2n+2),(2n+3)}$ and $H_{(2n+2),(2n+4)}$, $n = 0, 1,\ldots$ correspondingly.

In addition, Eq.~(\ref{eq:19}) explains the appearances of square root singularities \cite{PhysRev.184.1231} for non-Hermitian Hamiltonians. In the non-Hermitian case the expression under the square root can vanish, thus leading to branching points at $\epsilon = \epsilon_{\mathrm{branching\ point}}$. For $\epsilon< \epsilon_{\mathrm{branching\ point}}$ the energy levels coalesce and become complex conjugate pairs (see also Ref.~\cite{PhysRevA.95.052113}).
\begin{table}[t]
  \centering
  \begin{tabular}{| c  c  c  c  c  c  c  c  c  c|}
    \hline \hline
      $\epsilon$ & $n$ & $E_{\mathrm{exact}}$ & $E_{\mathrm{analytic}}$ & $E_{\mathrm{WKB}}$ & $\epsilon$ & $n$ & $E_{\mathrm{exact}}$ & $E_{\mathrm{analytic}}$ & $E_{\mathrm{WKB}}$\\
    \hline
    1 & 0 & 1.156  & 1.126  & 1.094  & 2 & 0 & 1.477  & 1.363  & 1.377 \\ 
      & 1 & 4.109  & 4.138  & 4.089  &   & 1 & 6.003  & 6.104  & 5.956 \\ 
      & 2 & 7.562  & 7.573  & 7.549  &   & 2 & 11.802 & 11.876 & 11.769\\
      & 3 & 11.314 & 11.290 & 11.304 &   & 3 & 18.459 & 18.417 & 18.432\\
      & 4 & 15.292 & 15.222 & 15.283 &   & 4 & 25.792 & 25.583 & 25.769\\
      & 5 & 19.452 & 19.332 & 19.444 &   & 5 & 33.694 & 33.284 & 33.675\\
      & 6 & 23.767 & 23.592 & 23.761 &   & 6 & 42.094 & 41.453 & 42.076\\
      & 7 & 28.176 & 27.985 & 28.212 &   & 7 & 50.937 & 50.044 & 50.921\\
      & 8 & 32.789 & 32.496 & 32.784 &   & 8 & 60.184 & 59.015 & 60.170\\
     \hline \hline
  \end{tabular}
  \caption{Comparison of the eigenvalues obtained through our analytic approximation $E_{\mathrm{analytic}}$ with the ones calculated numerically via the Arnoldi iteration $E_{\mathrm{exact}}$ and quasi-classical WKB approximation $E_{\mathrm{WKB}}$ \cite{0034-4885-70-6-R03}.}
  \label{tab:2}
\end{table}

\textit{Results and discussion.} --- In Fig.~\ref{fig:1} we compare the exact numerical solution with the one obtained by the implementation of the above mentioned steps 1-6. As follows from the figure the agreement between analytical and numerical solutions is remarkable. In Table~\ref{tab:2} we provide the numerical values of our analytic approximation, the exact numerical solution and the quasi-classical WKB approximation \cite{0034-4885-70-6-R03} for some selected values of $\epsilon$. We also remind that the quasi-classical WKB approximation was found only for $\epsilon>0$. In Fig.~\ref{fig:2} we demonstrate that after the branching point the energy levels coalesce and become complex conjugate pairs.

The results of our paper demonstrate that the suggested algorithm based on the solution of the system of Eqs.~(\ref{eq:3}), (\ref{eq:4}) is very effective for obtaining a good approximation for eigenvalues of $\mathcal{PT}$ symmetric Hamiltonians. Everything is reduced to the calculation of integrals, which define the matrix elements of the Hamiltonian operator. Consequently, our results could be especially important for quantum field theory and many-dimensional cases when the numerical solution of differential equations becomes problematic (in the Hermitian case please see Refs.~\cite{Feranchuk2015} and \cite{PhysRevD.92.125019,0305-4470-29-14-026,0022-3719-17-24-012}). It is also important to stress that the good approximation for the eigenvalues of localized states for the Hamiltonian (\ref{eq:1}) is mainly defined not by the exact asymptotic behavior of the wave functions $\sim \exp{(-a|x|^{(\epsilon+4)/2})}$ but rather by the correct positions around which they are localized together with the widths of their maxima. In our case these are the parameters $u$ and $\omega$ respectively.

Finally, since our zeroth-order approximation provides a very good accuracy we can assume that the iteration scheme of Refs.~\cite{Feranchuk2015,Feranchuk1995370} may be employed for the incorporation of corrections due to the off-diagonal matrix elements, thus offering a regular way to improve the zeroth-order approximation.

\textit{Outlook.} --- There is a current interest \cite{PhysRevA.95.052113} in the determination of the eigenvalues when $\epsilon < -1$. Consequently, we have tested the suggested approach for $-1.05 \le \epsilon < -1$. For this we also applied Eq.~(\ref{eq:19}), in which, however, we have mixed the odd and the even states in a different order. In this case we mixed $(0,1)$, $(2,3)$, ... states. The parameters $\omega_{n}$ and $u_{n}$ were chosen from the condition that the matrix elements $H_{n+1, n+2}$ and $H_{n+1, n+3}$, $n = 0, 1, \ldots$ are vanishing.

\begin{figure}[t]
  \centering
  \includegraphics[width=\columnwidth]{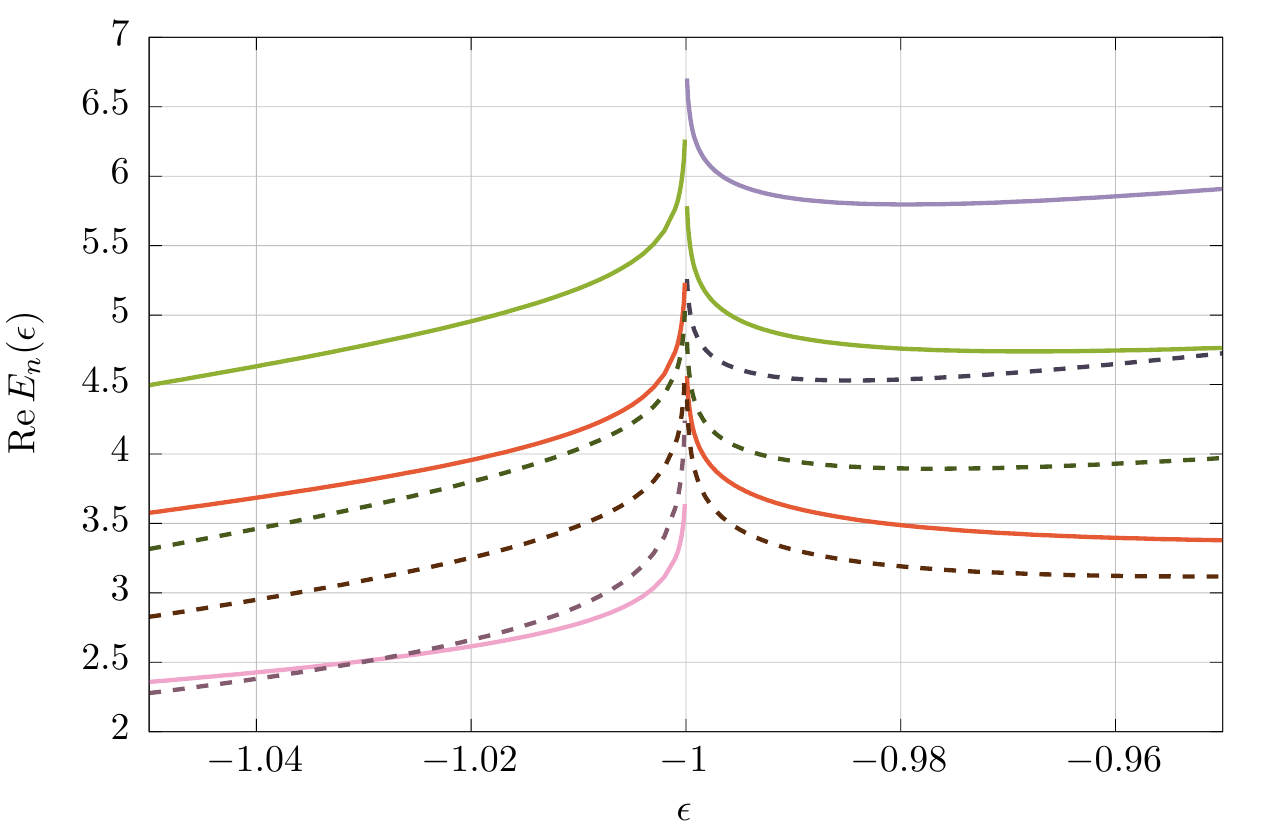}\\
  \includegraphics[width=\columnwidth]{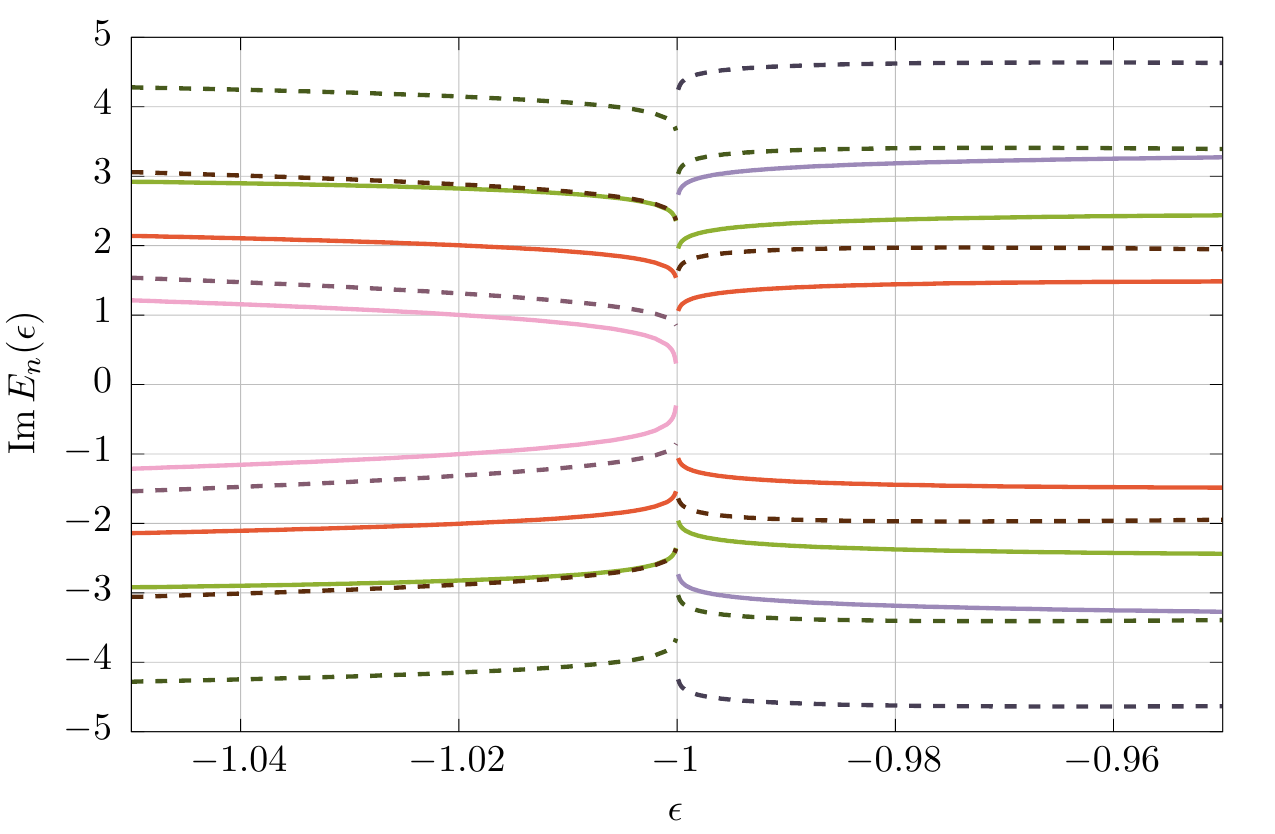}
  \caption{(Color online) The dependence of the real and imaginary parts of eigenvalues on the parameter $\epsilon$, in the region $-1.05 \le \epsilon < -0.95$. The dashed lines are the exact numerical solutions via the Arnoldi iteration. The solid lines are the analytic approximation via Eq.~(\ref{eq:19}) in which the states $(0,1)$, $(2,3)$, ... were mixed.}
  \label{fig:3}
\end{figure}

In Fig.~\ref{fig:3} we plot the dependence of the real and imaginary parts of eigenvalues on the parameter $\epsilon$, in the region $-1.05 \le \epsilon < -0.95$. By observing the figure we can conclude that all qualitative peculiarities of the system behavior are reproduced, however the accuracy of the results is somewhat worse, than for $\epsilon > -1$. Consequently, we suggest to employ the iteration scheme of Ref.~\cite{Feranchuk2015,Feranchuk1995370} in this case. In addition, one can try to introduce the third parameter $\alpha_{n}$ in the state function. Indeed, the choice of the basis functions $y_{n}(\alpha_{n}, \omega_{n}, u_{n},x) = \exp(\ri \alpha_{n}x)y_{n}(\omega_{n},u_{n},x)$ still allows to evaluate all matrix elements analytically, however, Eqs.~(\ref{eq:3}) and (\ref{eq:4}) should be generalized for the non-orthogonal basis \cite{PhysRevD.92.125019}. The latter is motivated by the $\mathcal{PT}$-symmetric definition of the scalar product, since in this case the phase of the wave function becomes important.

ODS is grateful to C. H. Keitel, S. M. Cavaletto, S. Bragin and A. Angioi for helpful discussions.

\bibliography{analytic_calc_ev_PTQM}

\begin{thebibliography}{70}%
\makeatletter
\providecommand \@ifxundefined [1]{%
 \@ifx{#1\undefined}
}%
\providecommand \@ifnum [1]{%
 \ifnum #1\expandafter \@firstoftwo
 \else \expandafter \@secondoftwo
 \fi
}%
\providecommand \@ifx [1]{%
 \ifx #1\expandafter \@firstoftwo
 \else \expandafter \@secondoftwo
 \fi
}%
\providecommand \natexlab [1]{#1}%
\providecommand \enquote  [1]{``#1''}%
\providecommand \bibnamefont  [1]{#1}%
\providecommand \bibfnamefont [1]{#1}%
\providecommand \citenamefont [1]{#1}%
\providecommand \href@noop [0]{\@secondoftwo}%
\providecommand \href [0]{\begingroup \@sanitize@url \@href}%
\providecommand \@href[1]{\@@startlink{#1}\@@href}%
\providecommand \@@href[1]{\endgroup#1\@@endlink}%
\providecommand \@sanitize@url [0]{\catcode `\\12\catcode `\$12\catcode
  `\&12\catcode `\#12\catcode `\^12\catcode `\_12\catcode `\%12\relax}%
\providecommand \@@startlink[1]{}%
\providecommand \@@endlink[0]{}%
\providecommand \url  [0]{\begingroup\@sanitize@url \@url }%
\providecommand \@url [1]{\endgroup\@href {#1}{\urlprefix }}%
\providecommand \urlprefix  [0]{URL }%
\providecommand \Eprint [0]{\href }%
\providecommand \doibase [0]{http://dx.doi.org/}%
\providecommand \selectlanguage [0]{\@gobble}%
\providecommand \bibinfo  [0]{\@secondoftwo}%
\providecommand \bibfield  [0]{\@secondoftwo}%
\providecommand \translation [1]{[#1]}%
\providecommand \BibitemOpen [0]{}%
\providecommand \bibitemStop [0]{}%
\providecommand \bibitemNoStop [0]{.\EOS\space}%
\providecommand \EOS [0]{\spacefactor3000\relax}%
\providecommand \BibitemShut  [1]{\csname bibitem#1\endcsname}%
\let\auto@bib@innerbib\@empty
\bibitem [{\citenamefont {Bender}\ and\ \citenamefont
  {Boettcher}(1998)}]{PhysRevLett.80.5243}%
  \BibitemOpen
  \bibfield  {author} {\bibinfo {author} {\bibfnamefont {C.~M.}\ \bibnamefont
  {Bender}}\ and\ \bibinfo {author} {\bibfnamefont {S.}~\bibnamefont
  {Boettcher}},\ }\href {\doibase 10.1103/PhysRevLett.80.5243} {\bibfield
  {journal} {\bibinfo  {journal} {Phys. Rev. Lett.}\ }\textbf {\bibinfo
  {volume} {80}},\ \bibinfo {pages} {5243} (\bibinfo {year}
  {1998})}\BibitemShut {NoStop}%
\bibitem [{\citenamefont {Bender}\ and\ \citenamefont
  {Turbiner}(1993)}]{BENDER1993442}%
  \BibitemOpen
  \bibfield  {author} {\bibinfo {author} {\bibfnamefont {C.~M.}\ \bibnamefont
  {Bender}}\ and\ \bibinfo {author} {\bibfnamefont {A.}~\bibnamefont
  {Turbiner}},\ }\href {\doibase 10.1016/0375-9601(93)90153-Q} {\bibfield
  {journal} {\bibinfo  {journal} {Physics Letters A}\ }\textbf {\bibinfo
  {volume} {173}},\ \bibinfo {pages} {442 } (\bibinfo {year}
  {1993})}\BibitemShut {NoStop}%
\bibitem [{\citenamefont {Bender}\ and\ \citenamefont
  {Wu}(1969)}]{PhysRev.184.1231}%
  \BibitemOpen
  \bibfield  {author} {\bibinfo {author} {\bibfnamefont {C.~M.}\ \bibnamefont
  {Bender}}\ and\ \bibinfo {author} {\bibfnamefont {T.~T.}\ \bibnamefont
  {Wu}},\ }\href {\doibase 10.1103/PhysRev.184.1231} {\bibfield  {journal}
  {\bibinfo  {journal} {Phys. Rev.}\ }\textbf {\bibinfo {volume} {184}},\
  \bibinfo {pages} {1231} (\bibinfo {year} {1969})}\BibitemShut {NoStop}%
\bibitem [{\citenamefont {Bender}(2007)}]{0034-4885-70-6-R03}%
  \BibitemOpen
  \bibfield  {author} {\bibinfo {author} {\bibfnamefont {C.~M.}\ \bibnamefont
  {Bender}},\ }\href {http://stacks.iop.org/0034-4885/70/i=6/a=R03} {\bibfield
  {journal} {\bibinfo  {journal} {Reports on Progress in Physics}\ }\textbf
  {\bibinfo {volume} {70}},\ \bibinfo {pages} {947} (\bibinfo {year}
  {2007})}\BibitemShut {NoStop}%
\bibitem [{\citenamefont {Dorey}\ \emph {et~al.}(2007)\citenamefont {Dorey},
  \citenamefont {Dunning},\ and\ \citenamefont {Tateo}}]{1751-8121-40-32-R01}%
  \BibitemOpen
  \bibfield  {author} {\bibinfo {author} {\bibfnamefont {P.}~\bibnamefont
  {Dorey}}, \bibinfo {author} {\bibfnamefont {C.}~\bibnamefont {Dunning}}, \
  and\ \bibinfo {author} {\bibfnamefont {R.}~\bibnamefont {Tateo}},\ }\href
  {http://stacks.iop.org/1751-8121/40/i=32/a=R01} {\bibfield  {journal}
  {\bibinfo  {journal} {Journal of Physics A: Mathematical and Theoretical}\
  }\textbf {\bibinfo {volume} {40}},\ \bibinfo {pages} {R205} (\bibinfo {year}
  {2007})}\BibitemShut {NoStop}%
\bibitem [{\citenamefont {Dorey}\ \emph {et~al.}(2001)\citenamefont {Dorey},
  \citenamefont {Dunning},\ and\ \citenamefont {Tateo}}]{0305-4470-34-28-305}%
  \BibitemOpen
  \bibfield  {author} {\bibinfo {author} {\bibfnamefont {P.}~\bibnamefont
  {Dorey}}, \bibinfo {author} {\bibfnamefont {C.}~\bibnamefont {Dunning}}, \
  and\ \bibinfo {author} {\bibfnamefont {R.}~\bibnamefont {Tateo}},\ }\href
  {http://stacks.iop.org/0305-4470/34/i=28/a=305} {\bibfield  {journal}
  {\bibinfo  {journal} {Journal of Physics A: Mathematical and General}\
  }\textbf {\bibinfo {volume} {34}},\ \bibinfo {pages} {5679} (\bibinfo {year}
  {2001})}\BibitemShut {NoStop}%
\bibitem [{\citenamefont {Bagchi}\ and\ \citenamefont
  {Roychoudhury}(2000)}]{0305-4470-33-1-101}%
  \BibitemOpen
  \bibfield  {author} {\bibinfo {author} {\bibfnamefont {B.}~\bibnamefont
  {Bagchi}}\ and\ \bibinfo {author} {\bibfnamefont {R.}~\bibnamefont
  {Roychoudhury}},\ }\href {http://stacks.iop.org/0305-4470/33/i=1/a=101}
  {\bibfield  {journal} {\bibinfo  {journal} {Journal of Physics A:
  Mathematical and General}\ }\textbf {\bibinfo {volume} {33}},\ \bibinfo
  {pages} {L1} (\bibinfo {year} {2000})}\BibitemShut {NoStop}%
\bibitem [{\citenamefont {Delabaere}\ and\ \citenamefont
  {Trinh}(2000)}]{0305-4470-33-48-314}%
  \BibitemOpen
  \bibfield  {author} {\bibinfo {author} {\bibfnamefont {E.}~\bibnamefont
  {Delabaere}}\ and\ \bibinfo {author} {\bibfnamefont {D.~T.}\ \bibnamefont
  {Trinh}},\ }\href {http://stacks.iop.org/0305-4470/33/i=48/a=314} {\bibfield
  {journal} {\bibinfo  {journal} {Journal of Physics A: Mathematical and
  General}\ }\textbf {\bibinfo {volume} {33}},\ \bibinfo {pages} {8771}
  (\bibinfo {year} {2000})}\BibitemShut {NoStop}%
\bibitem [{\citenamefont {Znojil}(1999)}]{ZNOJIL1999108}%
  \BibitemOpen
  \bibfield  {author} {\bibinfo {author} {\bibfnamefont {M.}~\bibnamefont
  {Znojil}},\ }\href {\doibase 10.1016/S0375-9601(99)00805-1} {\bibfield
  {journal} {\bibinfo  {journal} {Physics Letters A}\ }\textbf {\bibinfo
  {volume} {264}},\ \bibinfo {pages} {108 } (\bibinfo {year}
  {1999})}\BibitemShut {NoStop}%
\bibitem [{\citenamefont {Fernández}\ \emph {et~al.}(1999)\citenamefont
  {Fernández}, \citenamefont {Guardiola}, \citenamefont {Ros},\ and\
  \citenamefont {Znojil}}]{0305-4470-32-17-303}%
  \BibitemOpen
  \bibfield  {author} {\bibinfo {author} {\bibfnamefont {F.~M.}\ \bibnamefont
  {Fernández}}, \bibinfo {author} {\bibfnamefont {R.}~\bibnamefont
  {Guardiola}}, \bibinfo {author} {\bibfnamefont {J.}~\bibnamefont {Ros}}, \
  and\ \bibinfo {author} {\bibfnamefont {M.}~\bibnamefont {Znojil}},\ }\href
  {http://stacks.iop.org/0305-4470/32/i=17/a=303} {\bibfield  {journal}
  {\bibinfo  {journal} {Journal of Physics A: Mathematical and General}\
  }\textbf {\bibinfo {volume} {32}},\ \bibinfo {pages} {3105} (\bibinfo {year}
  {1999})}\BibitemShut {NoStop}%
\bibitem [{\citenamefont {Fring}\ and\ \citenamefont
  {Frith}(2017)}]{Fring20172318}%
  \BibitemOpen
  \bibfield  {author} {\bibinfo {author} {\bibfnamefont {A.}~\bibnamefont
  {Fring}}\ and\ \bibinfo {author} {\bibfnamefont {T.}~\bibnamefont {Frith}},\
  }\href {\doibase 10.1016/j.physleta.2017.05.041} {\bibfield  {journal}
  {\bibinfo  {journal} {Physics Letters A}\ }\textbf {\bibinfo {volume}
  {381}},\ \bibinfo {pages} {2318 } (\bibinfo {year} {2017})}\BibitemShut
  {NoStop}%
\bibitem [{\citenamefont {Luo}\ \emph {et~al.}(2017)\citenamefont {Luo},
  \citenamefont {Yang}, \citenamefont {Zhang}, \citenamefont {Li},\ and\
  \citenamefont {Yu}}]{PhysRevA.95.052128}%
  \BibitemOpen
  \bibfield  {author} {\bibinfo {author} {\bibfnamefont {X.}~\bibnamefont
  {Luo}}, \bibinfo {author} {\bibfnamefont {B.}~\bibnamefont {Yang}}, \bibinfo
  {author} {\bibfnamefont {X.}~\bibnamefont {Zhang}}, \bibinfo {author}
  {\bibfnamefont {L.}~\bibnamefont {Li}}, \ and\ \bibinfo {author}
  {\bibfnamefont {X.}~\bibnamefont {Yu}},\ }\href {\doibase
  10.1103/PhysRevA.95.052128} {\bibfield  {journal} {\bibinfo  {journal} {Phys.
  Rev. A}\ }\textbf {\bibinfo {volume} {95}},\ \bibinfo {pages} {052128}
  (\bibinfo {year} {2017})}\BibitemShut {NoStop}%
\bibitem [{\citenamefont {Klett}\ \emph {et~al.}(2017)\citenamefont {Klett},
  \citenamefont {Cartarius}, \citenamefont {Dast}, \citenamefont {Main},\ and\
  \citenamefont {Wunner}}]{PhysRevA.95.053626}%
  \BibitemOpen
  \bibfield  {author} {\bibinfo {author} {\bibfnamefont {M.}~\bibnamefont
  {Klett}}, \bibinfo {author} {\bibfnamefont {H.}~\bibnamefont {Cartarius}},
  \bibinfo {author} {\bibfnamefont {D.}~\bibnamefont {Dast}}, \bibinfo {author}
  {\bibfnamefont {J.}~\bibnamefont {Main}}, \ and\ \bibinfo {author}
  {\bibfnamefont {G.}~\bibnamefont {Wunner}},\ }\href {\doibase
  10.1103/PhysRevA.95.053626} {\bibfield  {journal} {\bibinfo  {journal} {Phys.
  Rev. A}\ }\textbf {\bibinfo {volume} {95}},\ \bibinfo {pages} {053626}
  (\bibinfo {year} {2017})}\BibitemShut {NoStop}%
\bibitem [{\citenamefont {Schwarz}\ \emph {et~al.}(2017)\citenamefont
  {Schwarz}, \citenamefont {Cartarius}, \citenamefont {Musslimani},
  \citenamefont {Main},\ and\ \citenamefont {Wunner}}]{PhysRevA.95.053613}%
  \BibitemOpen
  \bibfield  {author} {\bibinfo {author} {\bibfnamefont {L.}~\bibnamefont
  {Schwarz}}, \bibinfo {author} {\bibfnamefont {H.}~\bibnamefont {Cartarius}},
  \bibinfo {author} {\bibfnamefont {Z.~H.}\ \bibnamefont {Musslimani}},
  \bibinfo {author} {\bibfnamefont {J.}~\bibnamefont {Main}}, \ and\ \bibinfo
  {author} {\bibfnamefont {G.}~\bibnamefont {Wunner}},\ }\href {\doibase
  10.1103/PhysRevA.95.053613} {\bibfield  {journal} {\bibinfo  {journal} {Phys.
  Rev. A}\ }\textbf {\bibinfo {volume} {95}},\ \bibinfo {pages} {053613}
  (\bibinfo {year} {2017})}\BibitemShut {NoStop}%
\bibitem [{\citenamefont {Ahmed}\ \emph {et~al.}(2017)\citenamefont {Ahmed},
  \citenamefont {Herrero}, \citenamefont {Botey},\ and\ \citenamefont
  {Staliunas}}]{PhysRevA.95.053830}%
  \BibitemOpen
  \bibfield  {author} {\bibinfo {author} {\bibfnamefont {W.~W.}\ \bibnamefont
  {Ahmed}}, \bibinfo {author} {\bibfnamefont {R.}~\bibnamefont {Herrero}},
  \bibinfo {author} {\bibfnamefont {M.}~\bibnamefont {Botey}}, \ and\ \bibinfo
  {author} {\bibfnamefont {K.}~\bibnamefont {Staliunas}},\ }\href {\doibase
  10.1103/PhysRevA.95.053830} {\bibfield  {journal} {\bibinfo  {journal} {Phys.
  Rev. A}\ }\textbf {\bibinfo {volume} {95}},\ \bibinfo {pages} {053830}
  (\bibinfo {year} {2017})}\BibitemShut {NoStop}%
\bibitem [{\citenamefont {Zhao}\ and\ \citenamefont
  {Lu}(2017)}]{PhysRevLett.118.056401}%
  \BibitemOpen
  \bibfield  {author} {\bibinfo {author} {\bibfnamefont {Y.~X.}\ \bibnamefont
  {Zhao}}\ and\ \bibinfo {author} {\bibfnamefont {Y.}~\bibnamefont {Lu}},\
  }\href {\doibase 10.1103/PhysRevLett.118.056401} {\bibfield  {journal}
  {\bibinfo  {journal} {Phys. Rev. Lett.}\ }\textbf {\bibinfo {volume} {118}},\
  \bibinfo {pages} {056401} (\bibinfo {year} {2017})}\BibitemShut {NoStop}%
\bibitem [{\citenamefont {Bender}\ \emph
  {et~al.}(2017{\natexlab{a}})\citenamefont {Bender}, \citenamefont {Brody},\
  and\ \citenamefont {M\"uller}}]{PhysRevLett.118.130201}%
  \BibitemOpen
  \bibfield  {author} {\bibinfo {author} {\bibfnamefont {C.~M.}\ \bibnamefont
  {Bender}}, \bibinfo {author} {\bibfnamefont {D.~C.}\ \bibnamefont {Brody}}, \
  and\ \bibinfo {author} {\bibfnamefont {M.~P.}\ \bibnamefont {M\"uller}},\
  }\href {\doibase 10.1103/PhysRevLett.118.130201} {\bibfield  {journal}
  {\bibinfo  {journal} {Phys. Rev. Lett.}\ }\textbf {\bibinfo {volume} {118}},\
  \bibinfo {pages} {130201} (\bibinfo {year} {2017}{\natexlab{a}})}\BibitemShut
  {NoStop}%
\bibitem [{\citenamefont {Suchkov}\ \emph {et~al.}(2016)\citenamefont
  {Suchkov}, \citenamefont {Fotsa-Ngaffo}, \citenamefont {Kenfack-Jiotsa},
  \citenamefont {Tikeng}, \citenamefont {Kofane}, \citenamefont {Kivshar},\
  and\ \citenamefont {Sukhorukov}}]{1367-2630-18-6-065005}%
  \BibitemOpen
  \bibfield  {author} {\bibinfo {author} {\bibfnamefont {S.~V.}\ \bibnamefont
  {Suchkov}}, \bibinfo {author} {\bibfnamefont {F.}~\bibnamefont
  {Fotsa-Ngaffo}}, \bibinfo {author} {\bibfnamefont {A.}~\bibnamefont
  {Kenfack-Jiotsa}}, \bibinfo {author} {\bibfnamefont {A.~D.}\ \bibnamefont
  {Tikeng}}, \bibinfo {author} {\bibfnamefont {T.~C.}\ \bibnamefont {Kofane}},
  \bibinfo {author} {\bibfnamefont {Y.~S.}\ \bibnamefont {Kivshar}}, \ and\
  \bibinfo {author} {\bibfnamefont {A.~A.}\ \bibnamefont {Sukhorukov}},\ }\href
  {http://stacks.iop.org/1367-2630/18/i=6/a=065005} {\bibfield  {journal}
  {\bibinfo  {journal} {New Journal of Physics}\ }\textbf {\bibinfo {volume}
  {18}},\ \bibinfo {pages} {065005} (\bibinfo {year} {2016})}\BibitemShut
  {NoStop}%
\bibitem [{\citenamefont {Bagarello}(2016)}]{1751-8121-49-21-215304}%
  \BibitemOpen
  \bibfield  {author} {\bibinfo {author} {\bibfnamefont {F.}~\bibnamefont
  {Bagarello}},\ }\href {http://stacks.iop.org/1751-8121/49/i=21/a=215304}
  {\bibfield  {journal} {\bibinfo  {journal} {Journal of Physics A:
  Mathematical and Theoretical}\ }\textbf {\bibinfo {volume} {49}},\ \bibinfo
  {pages} {215304} (\bibinfo {year} {2016})}\BibitemShut {NoStop}%
\bibitem [{\citenamefont {Miao}\ and\ \citenamefont {Xu}(2016)}]{MIAO20161805}%
  \BibitemOpen
  \bibfield  {author} {\bibinfo {author} {\bibfnamefont {Y.-G.}\ \bibnamefont
  {Miao}}\ and\ \bibinfo {author} {\bibfnamefont {Z.-M.}\ \bibnamefont {Xu}},\
  }\href {\doibase 10.1016/j.physleta.2016.03.035} {\bibfield  {journal}
  {\bibinfo  {journal} {Physics Letters A}\ }\textbf {\bibinfo {volume}
  {380}},\ \bibinfo {pages} {1805 } (\bibinfo {year} {2016})}\BibitemShut
  {NoStop}%
\bibitem [{\citenamefont {Nixon}\ and\ \citenamefont
  {Yang}(2016)}]{PhysRevA.93.031802}%
  \BibitemOpen
  \bibfield  {author} {\bibinfo {author} {\bibfnamefont {S.}~\bibnamefont
  {Nixon}}\ and\ \bibinfo {author} {\bibfnamefont {J.}~\bibnamefont {Yang}},\
  }\href {\doibase 10.1103/PhysRevA.93.031802} {\bibfield  {journal} {\bibinfo
  {journal} {Phys. Rev. A}\ }\textbf {\bibinfo {volume} {93}},\ \bibinfo
  {pages} {031802} (\bibinfo {year} {2016})}\BibitemShut {NoStop}%
\bibitem [{\citenamefont {Simeonov}\ and\ \citenamefont
  {Vitanov}(2016)}]{PhysRevA.93.012123}%
  \BibitemOpen
  \bibfield  {author} {\bibinfo {author} {\bibfnamefont {L.~S.}\ \bibnamefont
  {Simeonov}}\ and\ \bibinfo {author} {\bibfnamefont {N.~V.}\ \bibnamefont
  {Vitanov}},\ }\href {\doibase 10.1103/PhysRevA.93.012123} {\bibfield
  {journal} {\bibinfo  {journal} {Phys. Rev. A}\ }\textbf {\bibinfo {volume}
  {93}},\ \bibinfo {pages} {012123} (\bibinfo {year} {2016})}\BibitemShut
  {NoStop}%
\bibitem [{\citenamefont {Brody}(2016)}]{1751-8121-49-10-10LT03}%
  \BibitemOpen
  \bibfield  {author} {\bibinfo {author} {\bibfnamefont {D.~C.}\ \bibnamefont
  {Brody}},\ }\href {http://stacks.iop.org/1751-8121/49/i=10/a=10LT03}
  {\bibfield  {journal} {\bibinfo  {journal} {Journal of Physics A:
  Mathematical and Theoretical}\ }\textbf {\bibinfo {volume} {49}},\ \bibinfo
  {pages} {10LT03} (\bibinfo {year} {2016})}\BibitemShut {NoStop}%
\bibitem [{\citenamefont {Bender}\ \emph
  {et~al.}(2017{\natexlab{b}})\citenamefont {Bender}, \citenamefont {Ghatak},\
  and\ \citenamefont {Gianfreda}}]{1751-8121-50-3-035601}%
  \BibitemOpen
  \bibfield  {author} {\bibinfo {author} {\bibfnamefont {C.~M.}\ \bibnamefont
  {Bender}}, \bibinfo {author} {\bibfnamefont {A.}~\bibnamefont {Ghatak}}, \
  and\ \bibinfo {author} {\bibfnamefont {M.}~\bibnamefont {Gianfreda}},\ }\href
  {http://stacks.iop.org/1751-8121/50/i=3/a=035601} {\bibfield  {journal}
  {\bibinfo  {journal} {Journal of Physics A: Mathematical and Theoretical}\
  }\textbf {\bibinfo {volume} {50}},\ \bibinfo {pages} {035601} (\bibinfo
  {year} {2017}{\natexlab{b}})}\BibitemShut {NoStop}%
\bibitem [{\citenamefont {Bender}\ \emph {et~al.}(2016)\citenamefont {Bender},
  \citenamefont {Hook}, \citenamefont {Mavromatos},\ and\ \citenamefont
  {Sarkar}}]{1751-8121-49-45-45LT01}%
  \BibitemOpen
  \bibfield  {author} {\bibinfo {author} {\bibfnamefont {C.~M.}\ \bibnamefont
  {Bender}}, \bibinfo {author} {\bibfnamefont {D.~W.}\ \bibnamefont {Hook}},
  \bibinfo {author} {\bibfnamefont {N.~E.}\ \bibnamefont {Mavromatos}}, \ and\
  \bibinfo {author} {\bibfnamefont {S.}~\bibnamefont {Sarkar}},\ }\href
  {http://stacks.iop.org/1751-8121/49/i=45/a=45LT01} {\bibfield  {journal}
  {\bibinfo  {journal} {Journal of Physics A: Mathematical and Theoretical}\
  }\textbf {\bibinfo {volume} {49}},\ \bibinfo {pages} {45LT01} (\bibinfo
  {year} {2016})}\BibitemShut {NoStop}%
\bibitem [{\citenamefont {Ramezani}(2017)}]{PhysRevA.96.011802}%
  \BibitemOpen
  \bibfield  {author} {\bibinfo {author} {\bibfnamefont {H.}~\bibnamefont
  {Ramezani}},\ }\href {\doibase 10.1103/PhysRevA.96.011802} {\bibfield
  {journal} {\bibinfo  {journal} {Phys. Rev. A}\ }\textbf {\bibinfo {volume}
  {96}},\ \bibinfo {pages} {011802} (\bibinfo {year} {2017})}\BibitemShut
  {NoStop}%
\bibitem [{\citenamefont {Bender}\ \emph {et~al.}(2010)\citenamefont {Bender},
  \citenamefont {Hook}, \citenamefont {Meisinger},\ and\ \citenamefont
  {Wang}}]{PhysRevLett.104.061601}%
  \BibitemOpen
  \bibfield  {author} {\bibinfo {author} {\bibfnamefont {C.~M.}\ \bibnamefont
  {Bender}}, \bibinfo {author} {\bibfnamefont {D.~W.}\ \bibnamefont {Hook}},
  \bibinfo {author} {\bibfnamefont {P.~N.}\ \bibnamefont {Meisinger}}, \ and\
  \bibinfo {author} {\bibfnamefont {Q.-h.}\ \bibnamefont {Wang}},\ }\href
  {\doibase 10.1103/PhysRevLett.104.061601} {\bibfield  {journal} {\bibinfo
  {journal} {Phys. Rev. Lett.}\ }\textbf {\bibinfo {volume} {104}},\ \bibinfo
  {pages} {061601} (\bibinfo {year} {2010})}\BibitemShut {NoStop}%
\bibitem [{\citenamefont {Znojil}(2017)}]{PhysRevA.96.012127}%
  \BibitemOpen
  \bibfield  {author} {\bibinfo {author} {\bibfnamefont {M.}~\bibnamefont
  {Znojil}},\ }\href {\doibase 10.1103/PhysRevA.96.012127} {\bibfield
  {journal} {\bibinfo  {journal} {Phys. Rev. A}\ }\textbf {\bibinfo {volume}
  {96}},\ \bibinfo {pages} {012127} (\bibinfo {year} {2017})}\BibitemShut
  {NoStop}%
\bibitem [{\citenamefont {Torosov}\ and\ \citenamefont
  {Vitanov}(2017)}]{PhysRevA.96.013845}%
  \BibitemOpen
  \bibfield  {author} {\bibinfo {author} {\bibfnamefont {B.~T.}\ \bibnamefont
  {Torosov}}\ and\ \bibinfo {author} {\bibfnamefont {N.~V.}\ \bibnamefont
  {Vitanov}},\ }\href {\doibase 10.1103/PhysRevA.96.013845} {\bibfield
  {journal} {\bibinfo  {journal} {Phys. Rev. A}\ }\textbf {\bibinfo {volume}
  {96}},\ \bibinfo {pages} {013845} (\bibinfo {year} {2017})}\BibitemShut
  {NoStop}%
\bibitem [{\citenamefont {Midya}\ and\ \citenamefont
  {Konotop}(2017)}]{PhysRevLett.119.033905}%
  \BibitemOpen
  \bibfield  {author} {\bibinfo {author} {\bibfnamefont {B.}~\bibnamefont
  {Midya}}\ and\ \bibinfo {author} {\bibfnamefont {V.~V.}\ \bibnamefont
  {Konotop}},\ }\href {\doibase 10.1103/PhysRevLett.119.033905} {\bibfield
  {journal} {\bibinfo  {journal} {Phys. Rev. Lett.}\ }\textbf {\bibinfo
  {volume} {119}},\ \bibinfo {pages} {033905} (\bibinfo {year}
  {2017})}\BibitemShut {NoStop}%
\bibitem [{\citenamefont {Caliceti}\ \emph {et~al.}(2005)\citenamefont
  {Caliceti}, \citenamefont {Graffi},\ and\ \citenamefont
  {Sjöstrand}}]{0305-4470-38-1-013}%
  \BibitemOpen
  \bibfield  {author} {\bibinfo {author} {\bibfnamefont {E.}~\bibnamefont
  {Caliceti}}, \bibinfo {author} {\bibfnamefont {S.}~\bibnamefont {Graffi}}, \
  and\ \bibinfo {author} {\bibfnamefont {J.}~\bibnamefont {Sjöstrand}},\
  }\href {http://stacks.iop.org/0305-4470/38/i=1/a=013} {\bibfield  {journal}
  {\bibinfo  {journal} {Journal of Physics A: Mathematical and General}\
  }\textbf {\bibinfo {volume} {38}},\ \bibinfo {pages} {185} (\bibinfo {year}
  {2005})}\BibitemShut {NoStop}%
\bibitem [{\citenamefont {Klaiman}\ \emph {et~al.}(2008)\citenamefont
  {Klaiman}, \citenamefont {G\"unther},\ and\ \citenamefont
  {Moiseyev}}]{PhysRevLett.101.080402}%
  \BibitemOpen
  \bibfield  {author} {\bibinfo {author} {\bibfnamefont {S.}~\bibnamefont
  {Klaiman}}, \bibinfo {author} {\bibfnamefont {U.}~\bibnamefont {G\"unther}},
  \ and\ \bibinfo {author} {\bibfnamefont {N.}~\bibnamefont {Moiseyev}},\
  }\href {\doibase 10.1103/PhysRevLett.101.080402} {\bibfield  {journal}
  {\bibinfo  {journal} {Phys. Rev. Lett.}\ }\textbf {\bibinfo {volume} {101}},\
  \bibinfo {pages} {080402} (\bibinfo {year} {2008})}\BibitemShut {NoStop}%
\bibitem [{\citenamefont {Peng}\ \emph
  {et~al.}(2014{\natexlab{a}})\citenamefont {Peng}, \citenamefont {Ozdemir},
  \citenamefont {Lei}, \citenamefont {Monifi}, \citenamefont {Gianfreda},
  \citenamefont {Long}, \citenamefont {Fan}, \citenamefont {Nori},
  \citenamefont {Bender},\ and\ \citenamefont {Yang}}]{bo_2014_5_394}%
  \BibitemOpen
  \bibfield  {author} {\bibinfo {author} {\bibfnamefont {B.}~\bibnamefont
  {Peng}}, \bibinfo {author} {\bibfnamefont {S.~K.}\ \bibnamefont {Ozdemir}},
  \bibinfo {author} {\bibfnamefont {F.}~\bibnamefont {Lei}}, \bibinfo {author}
  {\bibfnamefont {F.}~\bibnamefont {Monifi}}, \bibinfo {author} {\bibfnamefont
  {M.}~\bibnamefont {Gianfreda}}, \bibinfo {author} {\bibfnamefont {G.~L.}\
  \bibnamefont {Long}}, \bibinfo {author} {\bibfnamefont {S.}~\bibnamefont
  {Fan}}, \bibinfo {author} {\bibfnamefont {F.}~\bibnamefont {Nori}}, \bibinfo
  {author} {\bibfnamefont {C.~M.}\ \bibnamefont {Bender}}, \ and\ \bibinfo
  {author} {\bibfnamefont {L.}~\bibnamefont {Yang}},\ }\href {\doibase
  10.1038/nphys2927} {\bibfield  {journal} {\bibinfo  {journal} {Nature
  Physics}\ }\textbf {\bibinfo {volume} {10}},\ \bibinfo {pages} {394}
  (\bibinfo {year} {2014}{\natexlab{a}})}\BibitemShut {NoStop}%
\bibitem [{\citenamefont {Xiao}\ \emph {et~al.}(2017)\citenamefont {Xiao},
  \citenamefont {Zhan}, \citenamefont {Bian}, \citenamefont {Wang},
  \citenamefont {Zhang}, \citenamefont {Wang}, \citenamefont {Li},
  \citenamefont {Mochizuki}, \citenamefont {Kim}, \citenamefont {Kawakami},
  \citenamefont {Yi}, \citenamefont {Obuse}, \citenamefont {Sanders},\ and\
  \citenamefont {Xue}}]{xiao2017observation}%
  \BibitemOpen
  \bibfield  {author} {\bibinfo {author} {\bibfnamefont {L.}~\bibnamefont
  {Xiao}}, \bibinfo {author} {\bibfnamefont {X.}~\bibnamefont {Zhan}}, \bibinfo
  {author} {\bibfnamefont {Z.~H.}\ \bibnamefont {Bian}}, \bibinfo {author}
  {\bibfnamefont {K.~K.}\ \bibnamefont {Wang}}, \bibinfo {author}
  {\bibfnamefont {X.}~\bibnamefont {Zhang}}, \bibinfo {author} {\bibfnamefont
  {X.~P.}\ \bibnamefont {Wang}}, \bibinfo {author} {\bibfnamefont
  {J.}~\bibnamefont {Li}}, \bibinfo {author} {\bibfnamefont {K.}~\bibnamefont
  {Mochizuki}}, \bibinfo {author} {\bibfnamefont {D.}~\bibnamefont {Kim}},
  \bibinfo {author} {\bibfnamefont {N.}~\bibnamefont {Kawakami}}, \bibinfo
  {author} {\bibfnamefont {W.}~\bibnamefont {Yi}}, \bibinfo {author}
  {\bibfnamefont {H.}~\bibnamefont {Obuse}}, \bibinfo {author} {\bibfnamefont
  {B.~C.}\ \bibnamefont {Sanders}}, \ and\ \bibinfo {author} {\bibfnamefont
  {P.}~\bibnamefont {Xue}},\ }\href {\doibase 10.1038/nphys4204} {\bibfield
  {journal} {\bibinfo  {journal} {Nature Physics}\ } (\bibinfo {year} {2017}),\
  10.1038/nphys4204}\BibitemShut {NoStop}%
\bibitem [{\citenamefont {Yuto}\ \emph {et~al.}(2017)\citenamefont {Yuto},
  \citenamefont {Shunsuke},\ and\ \citenamefont {Ueda}}]{yuto2017parity}%
  \BibitemOpen
  \bibfield  {author} {\bibinfo {author} {\bibfnamefont {A.}~\bibnamefont
  {Yuto}}, \bibinfo {author} {\bibfnamefont {F.}~\bibnamefont {Shunsuke}}, \
  and\ \bibinfo {author} {\bibfnamefont {M.}~\bibnamefont {Ueda}},\ }\href
  {\doibase 10.1038/ncomms15791} {\bibfield  {journal} {\bibinfo  {journal}
  {Nature Communications}\ }\textbf {\bibinfo {volume} {8}},\ \bibinfo {pages}
  {15791} (\bibinfo {year} {2017})}\BibitemShut {NoStop}%
\bibitem [{\citenamefont {Wimmer}\ \emph {et~al.}(2015)\citenamefont {Wimmer},
  \citenamefont {Regensburger}, \citenamefont {Miri}, \citenamefont {Bersch},
  \citenamefont {Christodoulides},\ and\ \citenamefont
  {Peschel}}]{wimmer2015observation}%
  \BibitemOpen
  \bibfield  {author} {\bibinfo {author} {\bibfnamefont {M.}~\bibnamefont
  {Wimmer}}, \bibinfo {author} {\bibfnamefont {A.}~\bibnamefont
  {Regensburger}}, \bibinfo {author} {\bibfnamefont {M.-A.}\ \bibnamefont
  {Miri}}, \bibinfo {author} {\bibfnamefont {C.}~\bibnamefont {Bersch}},
  \bibinfo {author} {\bibfnamefont {D.~N.}\ \bibnamefont {Christodoulides}}, \
  and\ \bibinfo {author} {\bibfnamefont {U.}~\bibnamefont {Peschel}},\ }\href
  {\doibase 10.1038/ncomms8782} {\bibfield  {journal} {\bibinfo  {journal}
  {Nature Communications}\ }\textbf {\bibinfo {volume} {6}},\ \bibinfo {pages}
  {7782} (\bibinfo {year} {2015})}\BibitemShut {NoStop}%
\bibitem [{\citenamefont {Bender}\ \emph {et~al.}(2013)\citenamefont {Bender},
  \citenamefont {Gianfreda}, \citenamefont {\"Ozdemir}, \citenamefont {Peng},\
  and\ \citenamefont {Yang}}]{PhysRevA.88.062111}%
  \BibitemOpen
  \bibfield  {author} {\bibinfo {author} {\bibfnamefont {C.~M.}\ \bibnamefont
  {Bender}}, \bibinfo {author} {\bibfnamefont {M.}~\bibnamefont {Gianfreda}},
  \bibinfo {author} {\bibfnamefont {i.~m. c.~K.}\ \bibnamefont {\"Ozdemir}},
  \bibinfo {author} {\bibfnamefont {B.}~\bibnamefont {Peng}}, \ and\ \bibinfo
  {author} {\bibfnamefont {L.}~\bibnamefont {Yang}},\ }\href {\doibase
  10.1103/PhysRevA.88.062111} {\bibfield  {journal} {\bibinfo  {journal} {Phys.
  Rev. A}\ }\textbf {\bibinfo {volume} {88}},\ \bibinfo {pages} {062111}
  (\bibinfo {year} {2013})}\BibitemShut {NoStop}%
\bibitem [{\citenamefont {Peng}\ \emph
  {et~al.}(2014{\natexlab{b}})\citenamefont {Peng}, \citenamefont
  {{\"O}zdemir}, \citenamefont {Rotter}, \citenamefont {Yilmaz}, \citenamefont
  {Liertzer}, \citenamefont {Monifi}, \citenamefont {Bender}, \citenamefont
  {Nori},\ and\ \citenamefont {Yang}}]{Peng328}%
  \BibitemOpen
  \bibfield  {author} {\bibinfo {author} {\bibfnamefont {B.}~\bibnamefont
  {Peng}}, \bibinfo {author} {\bibfnamefont {{\c S}.~K.}\ \bibnamefont
  {{\"O}zdemir}}, \bibinfo {author} {\bibfnamefont {S.}~\bibnamefont {Rotter}},
  \bibinfo {author} {\bibfnamefont {H.}~\bibnamefont {Yilmaz}}, \bibinfo
  {author} {\bibfnamefont {M.}~\bibnamefont {Liertzer}}, \bibinfo {author}
  {\bibfnamefont {F.}~\bibnamefont {Monifi}}, \bibinfo {author} {\bibfnamefont
  {C.~M.}\ \bibnamefont {Bender}}, \bibinfo {author} {\bibfnamefont
  {F.}~\bibnamefont {Nori}}, \ and\ \bibinfo {author} {\bibfnamefont
  {L.}~\bibnamefont {Yang}},\ }\href {\doibase 10.1126/science.1258004}
  {\bibfield  {journal} {\bibinfo  {journal} {Science}\ }\textbf {\bibinfo
  {volume} {346}},\ \bibinfo {pages} {328} (\bibinfo {year}
  {2014}{\natexlab{b}})}\BibitemShut {NoStop}%
\bibitem [{\citenamefont {Schindler}\ \emph {et~al.}(2011)\citenamefont
  {Schindler}, \citenamefont {Li}, \citenamefont {Zheng}, \citenamefont
  {Ellis},\ and\ \citenamefont {Kottos}}]{PhysRevA.84.040101}%
  \BibitemOpen
  \bibfield  {author} {\bibinfo {author} {\bibfnamefont {J.}~\bibnamefont
  {Schindler}}, \bibinfo {author} {\bibfnamefont {A.}~\bibnamefont {Li}},
  \bibinfo {author} {\bibfnamefont {M.~C.}\ \bibnamefont {Zheng}}, \bibinfo
  {author} {\bibfnamefont {F.~M.}\ \bibnamefont {Ellis}}, \ and\ \bibinfo
  {author} {\bibfnamefont {T.}~\bibnamefont {Kottos}},\ }\href {\doibase
  10.1103/PhysRevA.84.040101} {\bibfield  {journal} {\bibinfo  {journal} {Phys.
  Rev. A}\ }\textbf {\bibinfo {volume} {84}},\ \bibinfo {pages} {040101}
  (\bibinfo {year} {2011})}\BibitemShut {NoStop}%
\bibitem [{\citenamefont {Zheng}\ \emph {et~al.}(2013)\citenamefont {Zheng},
  \citenamefont {Hao},\ and\ \citenamefont {Long}}]{Zheng20120053}%
  \BibitemOpen
  \bibfield  {author} {\bibinfo {author} {\bibfnamefont {C.}~\bibnamefont
  {Zheng}}, \bibinfo {author} {\bibfnamefont {L.}~\bibnamefont {Hao}}, \ and\
  \bibinfo {author} {\bibfnamefont {G.~L.}\ \bibnamefont {Long}},\ }\href
  {\doibase 10.1098/rsta.2012.0053} {\bibfield  {journal} {\bibinfo  {journal}
  {Philosophical Transactions of the Royal Society of London A: Mathematical,
  Physical and Engineering Sciences}\ }\textbf {\bibinfo {volume} {371}}
  (\bibinfo {year} {2013}),\ 10.1098/rsta.2012.0053}\BibitemShut {NoStop}%
\bibitem [{\citenamefont {Chtchelkatchev}\ \emph {et~al.}(2012)\citenamefont
  {Chtchelkatchev}, \citenamefont {Golubov}, \citenamefont {Baturina},\ and\
  \citenamefont {Vinokur}}]{PhysRevLett.109.150405}%
  \BibitemOpen
  \bibfield  {author} {\bibinfo {author} {\bibfnamefont {N.~M.}\ \bibnamefont
  {Chtchelkatchev}}, \bibinfo {author} {\bibfnamefont {A.~A.}\ \bibnamefont
  {Golubov}}, \bibinfo {author} {\bibfnamefont {T.~I.}\ \bibnamefont
  {Baturina}}, \ and\ \bibinfo {author} {\bibfnamefont {V.~M.}\ \bibnamefont
  {Vinokur}},\ }\href {\doibase 10.1103/PhysRevLett.109.150405} {\bibfield
  {journal} {\bibinfo  {journal} {Phys. Rev. Lett.}\ }\textbf {\bibinfo
  {volume} {109}},\ \bibinfo {pages} {150405} (\bibinfo {year}
  {2012})}\BibitemShut {NoStop}%
\bibitem [{\citenamefont {Regensburger}\ \emph {et~al.}(2012)\citenamefont
  {Regensburger}, \citenamefont {Bersch}, \citenamefont {Miri}, \citenamefont
  {Onishchukov}, \citenamefont {Christodoulides},\ and\ \citenamefont
  {Peschel}}]{Regensburger488_167_2012}%
  \BibitemOpen
  \bibfield  {author} {\bibinfo {author} {\bibfnamefont {A.}~\bibnamefont
  {Regensburger}}, \bibinfo {author} {\bibfnamefont {C.}~\bibnamefont
  {Bersch}}, \bibinfo {author} {\bibfnamefont {M.-A.}\ \bibnamefont {Miri}},
  \bibinfo {author} {\bibfnamefont {G.}~\bibnamefont {Onishchukov}}, \bibinfo
  {author} {\bibfnamefont {D.~N.}\ \bibnamefont {Christodoulides}}, \ and\
  \bibinfo {author} {\bibfnamefont {U.}~\bibnamefont {Peschel}},\ }\href
  {\doibase 10.1038/nature11298} {\bibfield  {journal} {\bibinfo  {journal}
  {Nature}\ }\textbf {\bibinfo {volume} {488}},\ \bibinfo {pages} {167}
  (\bibinfo {year} {2012})}\BibitemShut {NoStop}%
\bibitem [{\citenamefont {Liertzer}\ \emph {et~al.}(2012)\citenamefont
  {Liertzer}, \citenamefont {Ge}, \citenamefont {Cerjan}, \citenamefont
  {Stone}, \citenamefont {T\"ureci},\ and\ \citenamefont
  {Rotter}}]{PhysRevLett.108.173901}%
  \BibitemOpen
  \bibfield  {author} {\bibinfo {author} {\bibfnamefont {M.}~\bibnamefont
  {Liertzer}}, \bibinfo {author} {\bibfnamefont {L.}~\bibnamefont {Ge}},
  \bibinfo {author} {\bibfnamefont {A.}~\bibnamefont {Cerjan}}, \bibinfo
  {author} {\bibfnamefont {A.~D.}\ \bibnamefont {Stone}}, \bibinfo {author}
  {\bibfnamefont {H.~E.}\ \bibnamefont {T\"ureci}}, \ and\ \bibinfo {author}
  {\bibfnamefont {S.}~\bibnamefont {Rotter}},\ }\href {\doibase
  10.1103/PhysRevLett.108.173901} {\bibfield  {journal} {\bibinfo  {journal}
  {Phys. Rev. Lett.}\ }\textbf {\bibinfo {volume} {108}},\ \bibinfo {pages}
  {173901} (\bibinfo {year} {2012})}\BibitemShut {NoStop}%
\bibitem [{\citenamefont {Szameit}\ \emph {et~al.}(2011)\citenamefont
  {Szameit}, \citenamefont {Rechtsman}, \citenamefont {Bahat-Treidel},\ and\
  \citenamefont {Segev}}]{PhysRevA.84.021806}%
  \BibitemOpen
  \bibfield  {author} {\bibinfo {author} {\bibfnamefont {A.}~\bibnamefont
  {Szameit}}, \bibinfo {author} {\bibfnamefont {M.~C.}\ \bibnamefont
  {Rechtsman}}, \bibinfo {author} {\bibfnamefont {O.}~\bibnamefont
  {Bahat-Treidel}}, \ and\ \bibinfo {author} {\bibfnamefont {M.}~\bibnamefont
  {Segev}},\ }\href {\doibase 10.1103/PhysRevA.84.021806} {\bibfield  {journal}
  {\bibinfo  {journal} {Phys. Rev. A}\ }\textbf {\bibinfo {volume} {84}},\
  \bibinfo {pages} {021806} (\bibinfo {year} {2011})}\BibitemShut {NoStop}%
\bibitem [{\citenamefont {Chong}\ \emph {et~al.}(2011)\citenamefont {Chong},
  \citenamefont {Ge},\ and\ \citenamefont {Stone}}]{PhysRevLett.106.093902}%
  \BibitemOpen
  \bibfield  {author} {\bibinfo {author} {\bibfnamefont {Y.~D.}\ \bibnamefont
  {Chong}}, \bibinfo {author} {\bibfnamefont {L.}~\bibnamefont {Ge}}, \ and\
  \bibinfo {author} {\bibfnamefont {A.~D.}\ \bibnamefont {Stone}},\ }\href
  {\doibase 10.1103/PhysRevLett.106.093902} {\bibfield  {journal} {\bibinfo
  {journal} {Phys. Rev. Lett.}\ }\textbf {\bibinfo {volume} {106}},\ \bibinfo
  {pages} {093902} (\bibinfo {year} {2011})}\BibitemShut {NoStop}%
\bibitem [{\citenamefont {Bittner}\ \emph {et~al.}(2012)\citenamefont
  {Bittner}, \citenamefont {Dietz}, \citenamefont {G\"unther}, \citenamefont
  {Harney}, \citenamefont {Miski-Oglu}, \citenamefont {Richter},\ and\
  \citenamefont {Sch\"afer}}]{PhysRevLett.108.024101}%
  \BibitemOpen
  \bibfield  {author} {\bibinfo {author} {\bibfnamefont {S.}~\bibnamefont
  {Bittner}}, \bibinfo {author} {\bibfnamefont {B.}~\bibnamefont {Dietz}},
  \bibinfo {author} {\bibfnamefont {U.}~\bibnamefont {G\"unther}}, \bibinfo
  {author} {\bibfnamefont {H.~L.}\ \bibnamefont {Harney}}, \bibinfo {author}
  {\bibfnamefont {M.}~\bibnamefont {Miski-Oglu}}, \bibinfo {author}
  {\bibfnamefont {A.}~\bibnamefont {Richter}}, \ and\ \bibinfo {author}
  {\bibfnamefont {F.}~\bibnamefont {Sch\"afer}},\ }\href {\doibase
  10.1103/PhysRevLett.108.024101} {\bibfield  {journal} {\bibinfo  {journal}
  {Phys. Rev. Lett.}\ }\textbf {\bibinfo {volume} {108}},\ \bibinfo {pages}
  {024101} (\bibinfo {year} {2012})}\BibitemShut {NoStop}%
\bibitem [{\citenamefont {Rubinstein}\ \emph {et~al.}(2007)\citenamefont
  {Rubinstein}, \citenamefont {Sternberg},\ and\ \citenamefont
  {Ma}}]{PhysRevLett.99.167003}%
  \BibitemOpen
  \bibfield  {author} {\bibinfo {author} {\bibfnamefont {J.}~\bibnamefont
  {Rubinstein}}, \bibinfo {author} {\bibfnamefont {P.}~\bibnamefont
  {Sternberg}}, \ and\ \bibinfo {author} {\bibfnamefont {Q.}~\bibnamefont
  {Ma}},\ }\href {\doibase 10.1103/PhysRevLett.99.167003} {\bibfield  {journal}
  {\bibinfo  {journal} {Phys. Rev. Lett.}\ }\textbf {\bibinfo {volume} {99}},\
  \bibinfo {pages} {167003} (\bibinfo {year} {2007})}\BibitemShut {NoStop}%
\bibitem [{\citenamefont {Feng}\ \emph {et~al.}(2011)\citenamefont {Feng},
  \citenamefont {Ayache}, \citenamefont {Huang}, \citenamefont {Xu},
  \citenamefont {Lu}, \citenamefont {Chen}, \citenamefont {Fainman},\ and\
  \citenamefont {Scherer}}]{Feng729}%
  \BibitemOpen
  \bibfield  {author} {\bibinfo {author} {\bibfnamefont {L.}~\bibnamefont
  {Feng}}, \bibinfo {author} {\bibfnamefont {M.}~\bibnamefont {Ayache}},
  \bibinfo {author} {\bibfnamefont {J.}~\bibnamefont {Huang}}, \bibinfo
  {author} {\bibfnamefont {Y.-L.}\ \bibnamefont {Xu}}, \bibinfo {author}
  {\bibfnamefont {M.-H.}\ \bibnamefont {Lu}}, \bibinfo {author} {\bibfnamefont
  {Y.-F.}\ \bibnamefont {Chen}}, \bibinfo {author} {\bibfnamefont
  {Y.}~\bibnamefont {Fainman}}, \ and\ \bibinfo {author} {\bibfnamefont
  {A.}~\bibnamefont {Scherer}},\ }\href {\doibase 10.1126/science.1206038}
  {\bibfield  {journal} {\bibinfo  {journal} {Science}\ }\textbf {\bibinfo
  {volume} {333}},\ \bibinfo {pages} {729} (\bibinfo {year}
  {2011})}\BibitemShut {NoStop}%
\bibitem [{\citenamefont {Zhao}\ \emph {et~al.}(2010)\citenamefont {Zhao},
  \citenamefont {Schaden},\ and\ \citenamefont {Wu}}]{PhysRevA.81.042903}%
  \BibitemOpen
  \bibfield  {author} {\bibinfo {author} {\bibfnamefont {K.~F.}\ \bibnamefont
  {Zhao}}, \bibinfo {author} {\bibfnamefont {M.}~\bibnamefont {Schaden}}, \
  and\ \bibinfo {author} {\bibfnamefont {Z.}~\bibnamefont {Wu}},\ }\href
  {\doibase 10.1103/PhysRevA.81.042903} {\bibfield  {journal} {\bibinfo
  {journal} {Phys. Rev. A}\ }\textbf {\bibinfo {volume} {81}},\ \bibinfo
  {pages} {042903} (\bibinfo {year} {2010})}\BibitemShut {NoStop}%
\bibitem [{\citenamefont {Ruter}\ \emph {et~al.}(2010)\citenamefont {Ruter},
  \citenamefont {Makris}, \citenamefont {El-Ganainy}, \citenamefont
  {Christodoulides}, \citenamefont {Segev},\ and\ \citenamefont
  {Kip}}]{Ruter_Nature_6_192_2010}%
  \BibitemOpen
  \bibfield  {author} {\bibinfo {author} {\bibfnamefont {C.~E.}\ \bibnamefont
  {Ruter}}, \bibinfo {author} {\bibfnamefont {K.~G.}\ \bibnamefont {Makris}},
  \bibinfo {author} {\bibfnamefont {R.}~\bibnamefont {El-Ganainy}}, \bibinfo
  {author} {\bibfnamefont {D.~N.}\ \bibnamefont {Christodoulides}}, \bibinfo
  {author} {\bibfnamefont {M.}~\bibnamefont {Segev}}, \ and\ \bibinfo {author}
  {\bibfnamefont {D.}~\bibnamefont {Kip}},\ }\href {\doibase 10.1038/NPHYS1515}
  {\bibfield  {journal} {\bibinfo  {journal} {Nature Physics}\ }\textbf
  {\bibinfo {volume} {6}},\ \bibinfo {pages} {192} (\bibinfo {year}
  {2010})}\BibitemShut {NoStop}%
\bibitem [{\citenamefont {Bender}\ \emph
  {et~al.}(1999{\natexlab{a}})\citenamefont {Bender}, \citenamefont
  {Boettcher},\ and\ \citenamefont {Meisinger}}]{doi:10.1063/1.532860}%
  \BibitemOpen
  \bibfield  {author} {\bibinfo {author} {\bibfnamefont {C.~M.}\ \bibnamefont
  {Bender}}, \bibinfo {author} {\bibfnamefont {S.}~\bibnamefont {Boettcher}}, \
  and\ \bibinfo {author} {\bibfnamefont {P.~N.}\ \bibnamefont {Meisinger}},\
  }\href {\doibase 10.1063/1.532860} {\bibfield  {journal} {\bibinfo  {journal}
  {Journal of Mathematical Physics}\ }\textbf {\bibinfo {volume} {40}},\
  \bibinfo {pages} {2201} (\bibinfo {year} {1999}{\natexlab{a}})},\ \Eprint
  {http://arxiv.org/abs/http://dx.doi.org/10.1063/1.532860}
  {http://dx.doi.org/10.1063/1.532860} \BibitemShut {NoStop}%
\bibitem [{\citenamefont {Bender}\ \emph
  {et~al.}(2017{\natexlab{c}})\citenamefont {Bender}, \citenamefont
  {Hassanpour}, \citenamefont {Hook}, \citenamefont {Klevansky}, \citenamefont
  {S\"underhauf},\ and\ \citenamefont {Wen}}]{PhysRevA.95.052113}%
  \BibitemOpen
  \bibfield  {author} {\bibinfo {author} {\bibfnamefont {C.~M.}\ \bibnamefont
  {Bender}}, \bibinfo {author} {\bibfnamefont {N.}~\bibnamefont {Hassanpour}},
  \bibinfo {author} {\bibfnamefont {D.~W.}\ \bibnamefont {Hook}}, \bibinfo
  {author} {\bibfnamefont {S.~P.}\ \bibnamefont {Klevansky}}, \bibinfo {author}
  {\bibfnamefont {C.}~\bibnamefont {S\"underhauf}}, \ and\ \bibinfo {author}
  {\bibfnamefont {Z.}~\bibnamefont {Wen}},\ }\href {\doibase
  10.1103/PhysRevA.95.052113} {\bibfield  {journal} {\bibinfo  {journal} {Phys.
  Rev. A}\ }\textbf {\bibinfo {volume} {95}},\ \bibinfo {pages} {052113}
  (\bibinfo {year} {2017}{\natexlab{c}})}\BibitemShut {NoStop}%
\bibitem [{\citenamefont {Bender}\ \emph
  {et~al.}(1999{\natexlab{b}})\citenamefont {Bender}, \citenamefont {Cooper},
  \citenamefont {Meisinger},\ and\ \citenamefont {Savage}}]{BENDER1999224}%
  \BibitemOpen
  \bibfield  {author} {\bibinfo {author} {\bibfnamefont {C.~M.}\ \bibnamefont
  {Bender}}, \bibinfo {author} {\bibfnamefont {F.}~\bibnamefont {Cooper}},
  \bibinfo {author} {\bibfnamefont {P.~N.}\ \bibnamefont {Meisinger}}, \ and\
  \bibinfo {author} {\bibfnamefont {V.~M.}\ \bibnamefont {Savage}},\ }\href
  {\doibase 10.1016/S0375-9601(99)00468-5} {\bibfield  {journal} {\bibinfo
  {journal} {Physics Letters A}\ }\textbf {\bibinfo {volume} {259}},\ \bibinfo
  {pages} {224 } (\bibinfo {year} {1999}{\natexlab{b}})}\BibitemShut {NoStop}%
\bibitem [{\citenamefont {Feranchuk}\ \emph {et~al.}(2015)\citenamefont
  {Feranchuk}, \citenamefont {Ivanov}, \citenamefont {Le},\ and\ \citenamefont
  {Ulyanenkov}}]{Feranchuk2015}%
  \BibitemOpen
  \bibfield  {author} {\bibinfo {author} {\bibfnamefont {I.~D.}\ \bibnamefont
  {Feranchuk}}, \bibinfo {author} {\bibfnamefont {A.}~\bibnamefont {Ivanov}},
  \bibinfo {author} {\bibfnamefont {V.~H.}\ \bibnamefont {Le}}, \ and\ \bibinfo
  {author} {\bibfnamefont {A.~P.}\ \bibnamefont {Ulyanenkov}},\ }\href
  {https://books.google.de/books?id=uvWuoQEACAAJ} {\emph {\bibinfo {title}
  {Non-perturbative Description of Quantum Systems}}},\ Lecture Notes in
  Physics\ (\bibinfo  {publisher} {Springer International Publishing},\
  \bibinfo {year} {2015})\BibitemShut {NoStop}%
\bibitem [{\citenamefont {Feranchuk}\ \emph {et~al.}(1995)\citenamefont
  {Feranchuk}, \citenamefont {Komarov}, \citenamefont {Nichipor},\ and\
  \citenamefont {Ulyanenkov}}]{Feranchuk1995370}%
  \BibitemOpen
  \bibfield  {author} {\bibinfo {author} {\bibfnamefont {I.}~\bibnamefont
  {Feranchuk}}, \bibinfo {author} {\bibfnamefont {L.}~\bibnamefont {Komarov}},
  \bibinfo {author} {\bibfnamefont {I.}~\bibnamefont {Nichipor}}, \ and\
  \bibinfo {author} {\bibfnamefont {A.}~\bibnamefont {Ulyanenkov}},\ }\href
  {\doibase 10.1006/aphy.1995.1025} {\bibfield  {journal} {\bibinfo  {journal}
  {Annals of Physics}\ }\textbf {\bibinfo {volume} {238}},\ \bibinfo {pages}
  {370 } (\bibinfo {year} {1995})}\BibitemShut {NoStop}%
\bibitem [{\citenamefont {Trenogin}()}]{GalerkinMethod}%
  \BibitemOpen
  \bibfield  {author} {\bibinfo {author} {\bibfnamefont {V.~A.}\ \bibnamefont
  {Trenogin}},\ }\href@noop {} {\enquote {\bibinfo {title} {{Galerkin method.}
  {Encyclopedia of Mathematics.}}}\ }\bibinfo {howpublished}
  {\url{https://www.encyclopediaofmath.org/index.php/Galerkin_method}},\
  \bibinfo {note} {accessed: 13-09-2017}\BibitemShut {NoStop}%
\bibitem [{\citenamefont {Feranchuk}\ \emph {et~al.}(2016)\citenamefont
  {Feranchuk}, \citenamefont {Leonov},\ and\ \citenamefont
  {Skoromnik}}]{1751-8121-49-45-454001}%
  \BibitemOpen
  \bibfield  {author} {\bibinfo {author} {\bibfnamefont {I.~D.}\ \bibnamefont
  {Feranchuk}}, \bibinfo {author} {\bibfnamefont {A.~V.}\ \bibnamefont
  {Leonov}}, \ and\ \bibinfo {author} {\bibfnamefont {O.~D.}\ \bibnamefont
  {Skoromnik}},\ }\href {http://stacks.iop.org/1751-8121/49/i=45/a=454001}
  {\bibfield  {journal} {\bibinfo  {journal} {Journal of Physics A:
  Mathematical and Theoretical}\ }\textbf {\bibinfo {volume} {49}},\ \bibinfo
  {pages} {454001} (\bibinfo {year} {2016})}\BibitemShut {NoStop}%
\bibitem [{\citenamefont {Feranchuk}\ \emph {et~al.}(1996)\citenamefont
  {Feranchuk}, \citenamefont {Komarov},\ and\ \citenamefont
  {Ulyanenkov}}]{0305-4470-29-14-026}%
  \BibitemOpen
  \bibfield  {author} {\bibinfo {author} {\bibfnamefont {I.~D.}\ \bibnamefont
  {Feranchuk}}, \bibinfo {author} {\bibfnamefont {L.~I.}\ \bibnamefont
  {Komarov}}, \ and\ \bibinfo {author} {\bibfnamefont {A.~P.}\ \bibnamefont
  {Ulyanenkov}},\ }\href {http://stacks.iop.org/0305-4470/29/i=14/a=026}
  {\bibfield  {journal} {\bibinfo  {journal} {Journal of Physics A:
  Mathematical and General}\ }\textbf {\bibinfo {volume} {29}},\ \bibinfo
  {pages} {4035} (\bibinfo {year} {1996})}\BibitemShut {NoStop}%
\bibitem [{\citenamefont {Feranchuk}\ \emph {et~al.}(1984)\citenamefont
  {Feranchuk}, \citenamefont {Fisher},\ and\ \citenamefont
  {Komarov}}]{0022-3719-17-24-012}%
  \BibitemOpen
  \bibfield  {author} {\bibinfo {author} {\bibfnamefont {I.~D.}\ \bibnamefont
  {Feranchuk}}, \bibinfo {author} {\bibfnamefont {S.~I.}\ \bibnamefont
  {Fisher}}, \ and\ \bibinfo {author} {\bibfnamefont {L.~I.}\ \bibnamefont
  {Komarov}},\ }\href {http://stacks.iop.org/0022-3719/17/i=24/a=012}
  {\bibfield  {journal} {\bibinfo  {journal} {Journal of Physics C: Solid State
  Physics}\ }\textbf {\bibinfo {volume} {17}},\ \bibinfo {pages} {4309}
  (\bibinfo {year} {1984})}\BibitemShut {NoStop}%
\bibitem [{\citenamefont {Skoromnik}\ \emph {et~al.}(2015)\citenamefont
  {Skoromnik}, \citenamefont {Feranchuk}, \citenamefont {Lu},\ and\
  \citenamefont {Keitel}}]{PhysRevD.92.125019}%
  \BibitemOpen
  \bibfield  {author} {\bibinfo {author} {\bibfnamefont {O.~D.}\ \bibnamefont
  {Skoromnik}}, \bibinfo {author} {\bibfnamefont {I.~D.}\ \bibnamefont
  {Feranchuk}}, \bibinfo {author} {\bibfnamefont {D.~V.}\ \bibnamefont {Lu}}, \
  and\ \bibinfo {author} {\bibfnamefont {C.~H.}\ \bibnamefont {Keitel}},\
  }\href {\doibase 10.1103/PhysRevD.92.125019} {\bibfield  {journal} {\bibinfo
  {journal} {Phys. Rev. D}\ }\textbf {\bibinfo {volume} {92}},\ \bibinfo
  {pages} {125019} (\bibinfo {year} {2015})}\BibitemShut {NoStop}%
\bibitem [{\citenamefont {Skoromnik}\ \emph {et~al.}(2017)\citenamefont
  {Skoromnik}, \citenamefont {Feranchuk}, \citenamefont {Leonau},\ and\
  \citenamefont {Keitel}}]{1701.04800}%
  \BibitemOpen
  \bibfield  {author} {\bibinfo {author} {\bibfnamefont {O.~D.}\ \bibnamefont
  {Skoromnik}}, \bibinfo {author} {\bibfnamefont {I.~D.}\ \bibnamefont
  {Feranchuk}}, \bibinfo {author} {\bibfnamefont {A.~U.}\ \bibnamefont
  {Leonau}}, \ and\ \bibinfo {author} {\bibfnamefont {C.~H.}\ \bibnamefont
  {Keitel}},\ }\href {https://arxiv.org/abs/1701.04800} {\enquote {\bibinfo
  {title} {Analytic model of a multi-electron atom},}\ } (\bibinfo {year}
  {2017}),\ \Eprint {http://arxiv.org/abs/arXiv:1701.04800} {arXiv:1701.04800}
  \BibitemShut {NoStop}%
\bibitem [{\citenamefont {Moiseyev}(1998)}]{MOISEYEV1998212}%
  \BibitemOpen
  \bibfield  {author} {\bibinfo {author} {\bibfnamefont {N.}~\bibnamefont
  {Moiseyev}},\ }\href {\doibase 10.1016/S0370-1573(98)00002-7} {\bibfield
  {journal} {\bibinfo  {journal} {Physics Reports}\ }\textbf {\bibinfo {volume}
  {302}},\ \bibinfo {pages} {212 } (\bibinfo {year} {1998})}\BibitemShut
  {NoStop}%
\bibitem [{\citenamefont
  {Reinhardt}(1982)}]{doi:10.1146/annurev.pc.33.100182.001255}%
  \BibitemOpen
  \bibfield  {author} {\bibinfo {author} {\bibfnamefont {W.~P.}\ \bibnamefont
  {Reinhardt}},\ }\href {\doibase 10.1146/annurev.pc.33.100182.001255}
  {\bibfield  {journal} {\bibinfo  {journal} {Annual Review of Physical
  Chemistry}\ }\textbf {\bibinfo {volume} {33}},\ \bibinfo {pages} {223}
  (\bibinfo {year} {1982})}\BibitemShut {NoStop}%
\bibitem [{\citenamefont {Whittaker}\ and\ \citenamefont
  {Watson}(1996)}]{whittaker1996course}%
  \BibitemOpen
  \bibfield  {author} {\bibinfo {author} {\bibfnamefont {E.~T.}\ \bibnamefont
  {Whittaker}}\ and\ \bibinfo {author} {\bibfnamefont {G.~N.}\ \bibnamefont
  {Watson}},\ }\href {https://books.google.de/books?id=ULVdGZmi9VcC} {\emph
  {\bibinfo {title} {A Course of Modern Analysis}}},\ A Course of Modern
  Analysis: An Introduction to the General Theory of Infinite Processes and of
  Analytic Functions, with an Account of the Principal Transcendental
  Functions\ (\bibinfo  {publisher} {Cambridge University Press},\ \bibinfo
  {year} {1996})\BibitemShut {NoStop}%
\bibitem [{\citenamefont {Gradshteyn}\ and\ \citenamefont
  {Ryzhik}(2014)}]{gradshteyn2014table}%
  \BibitemOpen
  \bibfield  {author} {\bibinfo {author} {\bibfnamefont {I.~S.}\ \bibnamefont
  {Gradshteyn}}\ and\ \bibinfo {author} {\bibfnamefont {I.~M.}\ \bibnamefont
  {Ryzhik}},\ }\href {https://books.google.de/books?id=F7jiBQAAQBAJ} {\emph
  {\bibinfo {title} {Table of Integrals, Series, and Products}}}\ (\bibinfo
  {publisher} {Elsevier Science},\ \bibinfo {year} {2014})\BibitemShut
  {NoStop}%
\bibitem [{\citenamefont {Landau}\ and\ \citenamefont
  {Lifshitz}(1977)}]{LandauQM}%
  \BibitemOpen
  \bibfield  {author} {\bibinfo {author} {\bibfnamefont {L.~D.}\ \bibnamefont
  {Landau}}\ and\ \bibinfo {author} {\bibfnamefont {E.~M.}\ \bibnamefont
  {Lifshitz}},\ }\href {https://books.google.de/books?id=J9ui6KwC4mMC} {\emph
  {\bibinfo {title} {Quantum Mechanics: Non-relativistic Theory}}},\
  Butterworth Heinemann\ (\bibinfo  {publisher} {Butterworth-Heinemann},\
  \bibinfo {year} {1977})\BibitemShut {NoStop}%
\bibitem [{\citenamefont {Günther}\ and\ \citenamefont
  {Kirillov}(2006)}]{0305-4470-39-32-S08}%
  \BibitemOpen
  \bibfield  {author} {\bibinfo {author} {\bibfnamefont {U.}~\bibnamefont
  {Günther}}\ and\ \bibinfo {author} {\bibfnamefont {O.~N.}\ \bibnamefont
  {Kirillov}},\ }\href {http://stacks.iop.org/0305-4470/39/i=32/a=S08}
  {\bibfield  {journal} {\bibinfo  {journal} {Journal of Physics A:
  Mathematical and General}\ }\textbf {\bibinfo {volume} {39}},\ \bibinfo
  {pages} {10057} (\bibinfo {year} {2006})}\BibitemShut {NoStop}%
\bibitem [{\citenamefont {Graefe}\ \emph {et~al.}(2008)\citenamefont {Graefe},
  \citenamefont {Günther}, \citenamefont {Korsch},\ and\ \citenamefont
  {Niederle}}]{1751-8121-41-25-255206}%
  \BibitemOpen
  \bibfield  {author} {\bibinfo {author} {\bibfnamefont {E.~M.}\ \bibnamefont
  {Graefe}}, \bibinfo {author} {\bibfnamefont {U.}~\bibnamefont {Günther}},
  \bibinfo {author} {\bibfnamefont {H.~J.}\ \bibnamefont {Korsch}}, \ and\
  \bibinfo {author} {\bibfnamefont {A.~E.}\ \bibnamefont {Niederle}},\ }\href
  {http://stacks.iop.org/1751-8121/41/i=25/a=255206} {\bibfield  {journal}
  {\bibinfo  {journal} {Journal of Physics A: Mathematical and Theoretical}\
  }\textbf {\bibinfo {volume} {41}},\ \bibinfo {pages} {255206} (\bibinfo
  {year} {2008})}\BibitemShut {NoStop}%
\bibitem [{\citenamefont {Demange}\ and\ \citenamefont
  {Graefe}(2012)}]{1751-8121-45-2-025303}%
  \BibitemOpen
  \bibfield  {author} {\bibinfo {author} {\bibfnamefont {G.}~\bibnamefont
  {Demange}}\ and\ \bibinfo {author} {\bibfnamefont {E.-M.}\ \bibnamefont
  {Graefe}},\ }\href {http://stacks.iop.org/1751-8121/45/i=2/a=025303}
  {\bibfield  {journal} {\bibinfo  {journal} {Journal of Physics A:
  Mathematical and Theoretical}\ }\textbf {\bibinfo {volume} {45}},\ \bibinfo
  {pages} {025303} (\bibinfo {year} {2012})}\BibitemShut {NoStop}%
\bibitem [{\citenamefont {Heiss}\ and\ \citenamefont
  {Wunner}(2016)}]{1751-8121-49-49-495303}%
  \BibitemOpen
  \bibfield  {author} {\bibinfo {author} {\bibfnamefont {W.~D.}\ \bibnamefont
  {Heiss}}\ and\ \bibinfo {author} {\bibfnamefont {G.}~\bibnamefont {Wunner}},\
  }\href {http://stacks.iop.org/1751-8121/49/i=49/a=495303} {\bibfield
  {journal} {\bibinfo  {journal} {Journal of Physics A: Mathematical and
  Theoretical}\ }\textbf {\bibinfo {volume} {49}},\ \bibinfo {pages} {495303}
  (\bibinfo {year} {2016})}\BibitemShut {NoStop}%
\end{thebibliography}%

\end{document}